\documentclass[10pt,twocolumn,letterpaper]{article}

\usepackage[pagenumbers]{iccv} 

\usepackage{array}
\usepackage{caption}
\usepackage{subcaption}
\usepackage{stackengine}
\usepackage{wrapfig}
\usepackage{tablefootnote}
\usepackage{tabularx}

\usepackage{cuted}      
\usepackage{caption}   
\usepackage[export]{adjustbox}
\usepackage{booktabs}
\usepackage{multirow}
\usepackage{makecell}
\usepackage{algorithm}
\usepackage{algpseudocode}
\usepackage{url}
\usepackage{tikz}
\usetikzlibrary{calc,spy}
\usepackage{cancel}
\usepackage{diagbox}
\usepackage{pifont}
\usepackage[accsupp]{axessibility}
\usetikzlibrary{calc,spy}
\usepackage{cancel}
\usepackage{amsthm}
\usepackage{graphicx}
\usepackage{amsmath}
\usepackage{amssymb}
\usepackage{booktabs}

\usepackage{array}
\usepackage{caption}
\usepackage{subcaption}
\usepackage{stackengine}
\usepackage{wrapfig}

\usepackage[export]{adjustbox}
\usepackage{booktabs}
\usepackage{multirow}
\usepackage{makecell}
\usepackage{algorithm}
\usepackage{algpseudocode}
\usepackage{url}
\usepackage{tikz}
\usetikzlibrary{calc,spy}
\usepackage{cancel}
\usepackage{diagbox}
\usepackage{pifont}
\usepackage[accsupp]{axessibility}
\newcommand*\samethanks[1][\value{footnote}]{\footnotemark[#1]}

\definecolor{iccvblue}{rgb}{0.21,0.49,0.74}
\usepackage[pagebackref,breaklinks,colorlinks,allcolors=iccvblue]{hyperref}

\def\paperID{10926} 
\def\confName{ICCV}
\def\confYear{2025}

\title{JPEG Processing Neural Operator for Backward-Compatible Coding}

\author{
Woo Kyoung Han$^{1}$\thanks{Equal Contribution.}  \qquad\qquad\qquad Yongjun Lee$^1$\samethanks[1]  \qquad\qquad\qquad Byeonghun Lee$^1$ \\
  \quad Sang Hyun Park$^2$   \qquad\qquad\qquad Sunghoon Im$^2$\thanks{Corresponding author.}  \qquad\qquad\qquad Kyong Hwan Jin$^1$\samethanks[2]\\
$^1$Korea University \qquad\qquad\qquad $^2$DGIST\\
{\tt\footnotesize \{wookyoung0727, lyj9805, byeonghun\_lee, kyong\_jin\}@korea.ac.kr,} {\tt\footnotesize \{shpark13135, sunghoonim\}@dgist.ac.kr}
}

\begin{document}

\maketitle

\tikzstyle{largewindow_w} = [white, line width=0.30mm]
\tikzstyle{smallwindow_w} = [white, line width=0.10mm]
\tikzstyle{largewindow_b} = [blue, line width=0.30mm]
\tikzstyle{smallwindow_b} = [blue, line width=0.10mm]
\tikzstyle{closeup_b} = [
  opacity=1.0,          
  height=1cm,         
  width=1cm,          
  connect spies, blue  
]
\tikzstyle{closeup_w} = [
  opacity=1.0,          
  height=1cm,         
  width=1cm,          
  connect spies, white  
]
\tikzstyle{closeup_w_2} = [
  opacity=1.0,          
  height=1.7cm,         
  width=1.7cm,          
  connect spies, white  
]
\tikzstyle{closeup_w_3} = [
  opacity=1.0,          
  height=1.35cm,         
  width=1.35cm,          
  connect spies, white  
]
\tikzstyle{closeup_w_4} = [
  opacity=1.0,          
  height=0.8cm,         
  width=0.8cm,          
  connect spies, white  
]
\tikzstyle{closeup_w_5} = [
  opacity=1.0,          
  height=0.6in,         
  width=0.6in,          
  connect spies, white  
]

\begin{abstract}
Despite significant advances in learning-based lossy compression algorithms, standardizing codecs remains a critical challenge. 
In this paper, we present the JPEG Processing Neural Operator (JPNeO), a next-generation JPEG algorithm that maintains full backward compatibility with the current JPEG format. 
Our JPNeO improves chroma component preservation and enhances reconstruction fidelity compared to existing artifact removal methods by incorporating neural operators in both the encoding and decoding stages. 
JPNeO achieves practical benefits in terms of reduced memory usage and parameter count. 
We further validate our hypothesis about the existence of a space with high mutual information through empirical evidence.
In summary, the JPNeO functions as a high-performance out-of-the-box image compression pipeline without changing source coding's protocol.
Our source code is available at \url{https://github.com/WooKyoungHan/JPNeO}.

\end{abstract}
    
\section{Introduction}\label{sec:intro}

 All compression algorithms depend on how a signal is transformed into a symbol, also known as source coding \cite{shannoncapacity,informationtheorybook}. With the advancement of Deep Neural Networks (DNNs), lossy image compression has progressed towards more effective transformation \cite{balle2016end,balle2018variational,minnen2018joint}. DNN-based approaches utilizing nonlinear transform coding show remarkable performance in compression. Consequently, research on standardization is actively progressing in line with these advancements \cite{jpegai20021white, jpegai2020ascenso}.
 
However, a significant proportion of images have already been compressed using legacy codecs, and the standardization of DNN-based compression methods is expected to require considerable time. Moreover, among legacy codecs, JPEG \cite{jpegstandard} compression is widely integrated into image signal processors (ISPs) \cite{nakamura2017imageisp}, making it an inevitable part of the image processing pipeline. 
\begin{figure}
    \vspace{-5pt}
    \centering
    \includegraphics[trim=0 0 0 0 ,clip,width = 3.1in]{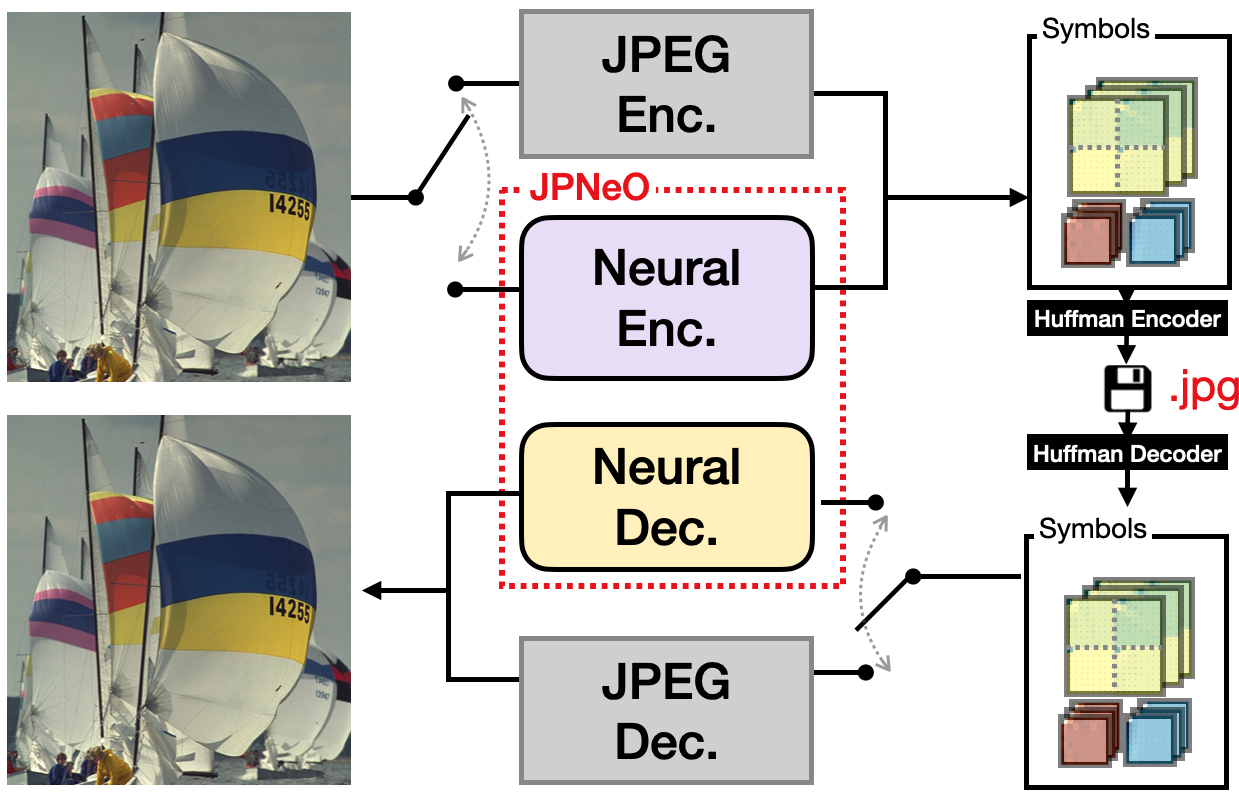}

    \vspace{-10pt}

\caption{\textbf{Overall Concept} of the proposed JPNeO method. Our method enables flexible switching between the conventional JPEG encoder and decoder as needed. Our approach ensures interoperability, allowing conventional JPEG-encoded files to be decoded with a conventional JPEG decoder and existing JPEG files to be decoded using our method.}
    \label{fig:concept}
   \vspace{-20pt}
\end{figure}
Therefore, research on DNN-based approaches for JPEG is actively ongoing with a focus on the compression artifact removal \cite{fbcnn,qgac,arcnn} and the exploration encoding images \cite{learnablejpeg,preeditingimage}. 
Several notable studies have explored processing JPEG images without relying on the conventional decoder. 
\citet{rgbnomore} bypass the JPEG decoder and apply the data to classification tasks.   
\citet{jdec2024han} bypass the JPEG decoder and propose a DNN decoder generating high-quality images. 

\begin{figure}[t]
\footnotesize
\centering
\includegraphics[width = 3.1in]{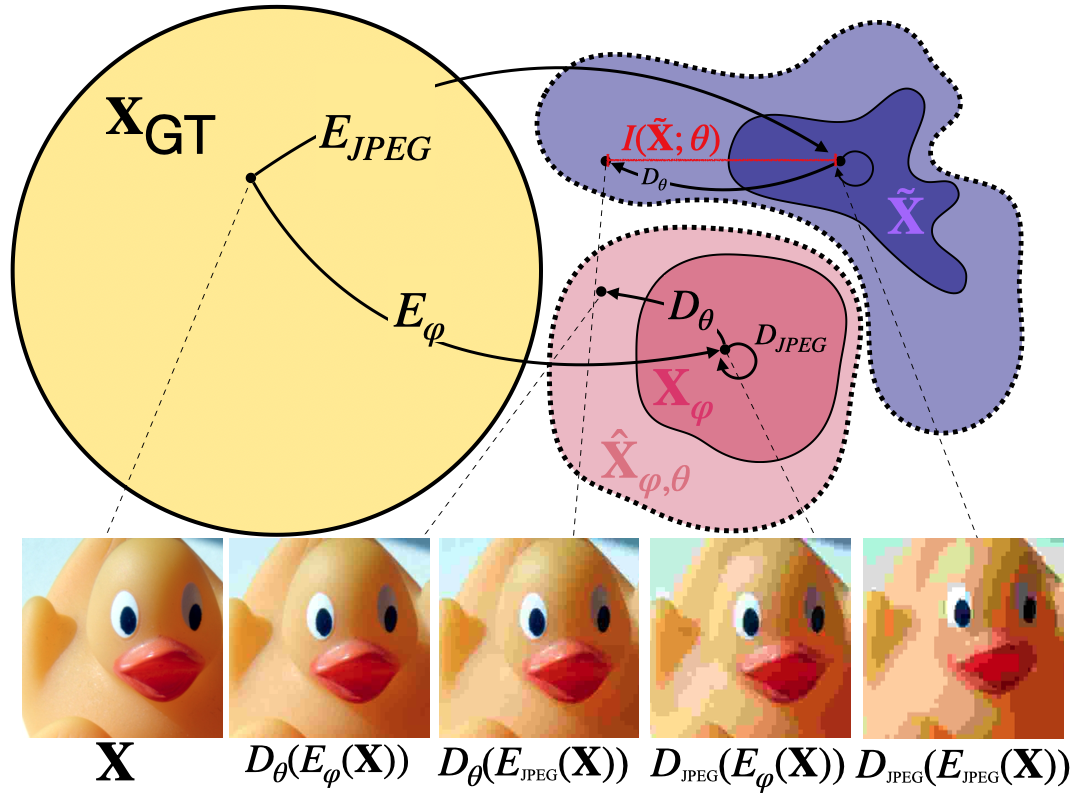}
\vspace{-10pt}
\caption{\textbf{Illustration of image sets under the JPEG process.} $E_{\varphi/\text{JPEG}}$ and $D_{\theta/\text{JPEG}}$ indicate the neural/JPEG encoder and decoder, respectively. $E_\varphi$ maps images to a range with lower distortion than the $E_\text{JPEG}$ range. $D_\text{JPEG}$ is limited to its original range. Our JPNeO's $D_{\theta}$ directs images to a range with lower distortion.}
\vspace*{-10pt}
\label{fig:jpeg_mapping}
\vspace{-10pt}
\end{figure}

Building on these advancements, we propose a complete \textbf{J}PEG \textbf{P}rocessing \textbf{Ne}ural \textbf{O}perator (\textbf{JPNeO}), DNN-based codec that is fully backward-compatible with the existing JPEG. \cref{fig:concept} presents the overall concept of our JPNeO. Our JPNeO consists of a JPEG encoding and decoding neural operator (JENO and JDNO), allowing flexible replacement with conventional JPEG encoders and decoders. Conventional JPEG decoders rely solely on encoder-constrained information. To overcome this limitation, we integrate a decoder that learns image priors enabling high-quality image reconstruction. To improve the JPEG encoder, we extend existing research \cite{learnablejpeg} to develop an encoder that maps images to a space optimized for JPEG. We hypothesise that a more efficient latent space exists where the latent vectors are relatively closer to the ground truth. As demonstrated in \cref{fig:jpeg_mapping}, the embedded mutual information of the JPNeO is leveraged for both encoding and decoding high-quality images. We provide experimental validation to support this interpretation. By incorporating a neural operator in the encoding and decoding stages, we propose a novel backward-compatible JPEG processing model that is efficient in parameters and memory usage.


In summary, our main contributions are as follows:
\begin{itemize}
    \item We propose a learning-based codec, JPNeO, that is fully compatible with the existing JPEG storage format in all cases, ultimately achieving state-of-the-art performance within the JPEG compression framework.
    
    \item We provide experimental validation to demonstrate that our encoder and decoder increase the mutual information with the original image at high and low bit rates respectively.

    \item Our approach is efficient in terms of memory usage and parameter count, and this efficiency is further enhanced by the flexibility to decouple the encoder and decoder.

\end{itemize}

\section{Related Work}\label{sec:related}

\textbf{Lossy Image Compression}
Lossy compression is a core technology that reduces file size by discarding information while preserving perceptual quality and methods like JPEG \cite{jpegstandard} and JPEG2000 \cite{taubman2002jpeg2000} have been industry standards based on transform coding. More recently, re‑compression techniques—including JPEG XL \cite{alakuijala2019jpegxl} and the learned lossless JPEG re-compression of \citet{guo2022jpegrecompression}—have been proposed to achieve higher efficiency while remaining practical for real‑world deployment.
Although DNN-based methods with non-linear transform \cite{balle2016end,balle2018variational,minnen2018joint} show remarkable performance in lossy compression, the widespread adoption of legacy standards makes it challenging to replace existing standards and image processing pipelines. Therefore, it is crucial to develop methods that maintain backward compatibility with existing standards. In this paper, we focus on JPEG\cite{jpegstandard}, one of the most widely used compression standards. The proposed JPNeO is designed to achieve full backward compatibility by enabling selective use of either neural or legacy components in both the encoder and decoder, thereby allowing seamless operation with existing JPEG bitstreams.



\noindent{\bf Deep Learning for JPEG} 
Following the advancement of DNNs, image restoration methods \cite{srcnn, cas-cnn,jpegrelateddedicated,arcnn,jpegrelatedwaveletcnn, bahat2021whatsin,jdec2024han, fbcnn, qgac} have been proposed to mitigate the distortion introduced during the encoding process. 
JPEG artifact removal DNNs work as auxiliary decoders integrated into the JPEG decoder. 
\citet{qgac} utilize a JPEG-aware approach that takes the quantization matrix as a prior and restores luma and chroma components.
\citet{fbcnn} proposed a method to restore JPEG images with a blind quality factor, aiming to address the common problem of multiple JPEG compression in real-world scenarios. 
Recently, \citet{jdec2024han} proposed an approach that decodes images from spectra, bypassing the conventional JPEG decoding process.
In addition to decoding strategies, encoding-oriented approaches have also been explored. \citet{preeditingimage} proposed a method that pre-edits images for better compression. \citet{learnablejpeg} suggested a trainable quantization matrix and merge with the pre-editing method. 
Previous studies have primarily focused on enhancing the encoder or decoder respectively and each module operates independently without sharing information. Building on these advancements, we have developed a fully JPEG-compatible codec that takes advantage of handling mutual information. 



\begin{figure*}[t]
\centering
\includegraphics[trim= 0 0 0 0,clip,width=7in]{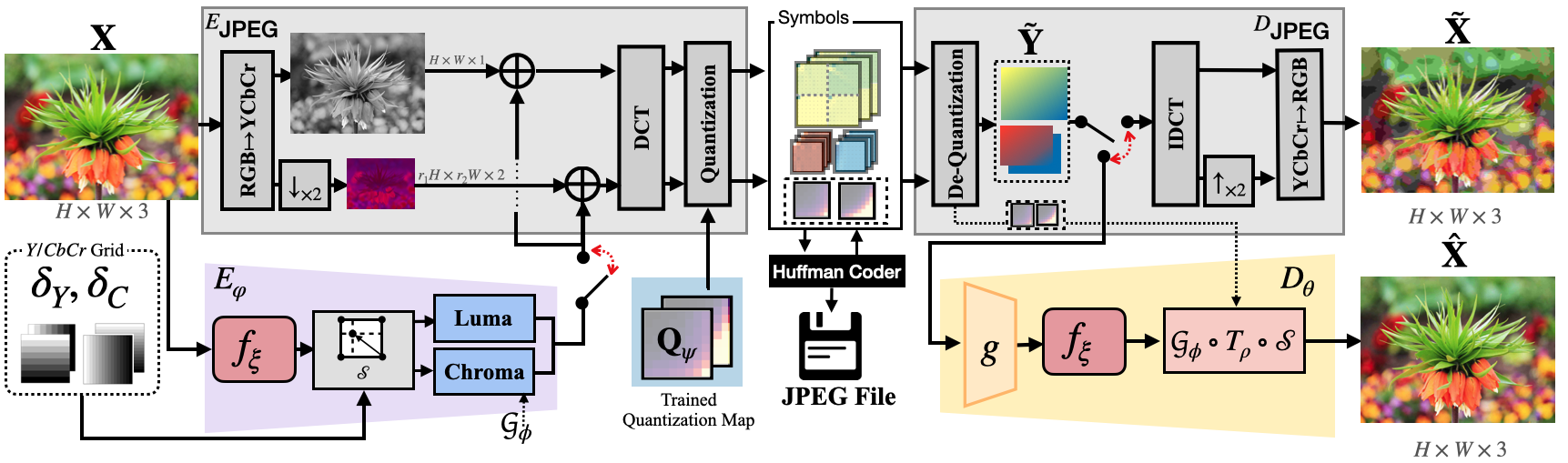}
\vspace{-20pt}
\caption{\textbf{Schematic overview of our JPNeO.} JPNeO consists of a JPEG  encoding neral operator (JENO ($E_\varphi$)), a pre-trained quantization matrix ($\mathbf{Q}_\psi$), and a JPEG decoding neural operator (JDNO ($D_\theta$)). Our JENO, consisting of a feature extractor ($f_\xi$), sampling ($\mathcal{S}$), and neural operators ($\mathcal{G}_\phi$) functions as an auxiliary encoder to the existing JPEG encoder ($E_\text{JPEG}$). JDNO, conposed of a group embedding ($g$), feature extractor $f_\xi$, and Cosine Neural Opearator ($\mathcal{G}_\phi\circ \mathbf{T}_\rho \circ \mathcal{S}$) can replace the existing JPEG decoder ($D_\text{JPEG}$). Ultimately, our JPNeO is designed to function properly even if each module is selectively applied.}
\label{Figmain}
\vspace{-12pt}
\end{figure*}

\noindent{\bf Neural Operator} Several studies interpret an image as a 2D function, i.e., implicit neural representations (INR) \cite{sitzmann2019siren,lee2021local,lee2022learning,itsrn,chen2021learning,wirewavelet2023}. Additionally, \cite{lee2021local,chen2021learning,wirewavelet2023} applied INRs to inverse problems. For JPEG, \citet{jdec2024han} interpreted decoding spectra using an INR method. Recently, a novel neural operator method that addresses inverse problems by solving differential equations \cite{lu2021learning, gupta2021multiwavelet, wei2023super, li2020fourier, kovachki2023neural} has shown remarkable performance. The neural operator was designed to solve the differential equation below \cite{kovachki2023neural}:
\vspace{-7pt}
\begin{align}
    (\mathbf{L}_a u)(x) &= f(x), \quad x\in D    \label{eq:no_init} \\
    u(x) &=0, \quad x\in \partial D
\end{align}
where $\mathbf{L}:\mathcal{A}\mapsto \mathbf{L}(\mathcal{U;\mathcal{U}^*})$ indicates differential operator and $a\in\mathcal{A}$ is coefficient functions of the PDE. The neural operator $\mathcal{G}_\theta: \mathcal{A}\mapsto \mathcal{U}\ni u$ is an empirical solver for \cref{eq:no_init}.
 Previous studies \cite{li2020fourier, kovachki2023neural} have demonstrated that the Galerkin-type attention method is equivalent to the Galerkin method \cite{galerkin_ern2004theory} and acts as the solver.
We leverage this approach for both JPEG encoding and decoding to enhance our JPEG codec.

\section{Problem Formulation}\label{sec:problemformulation}

\textbf{Preliminary}  Let $\mathbf{X} \in \mathbb{R}^{H\times W \times 3}$ be a ground-truth image.  The JPEG \cite{jpegstandard} contains an encoder ($E_\text{JPEG}: \mathbf{X}\mapsto \mathbf{Y'}$) and a decoder ($D_\text{JPEG}: \mathbf{Y'}\mapsto \mathbf{\tilde X}$). A source coding of the JPEG is demonstrated by the equation below:
\vspace{-10pt}
\begin{align}\label{eq:jpeg_oneline}
      \overbrace{{\mathbf{X} \overset{\textcolor{red}{\downarrow_{\times 2}}}{\longrightarrow} \mathbf{X'}}  \overset{\text{DCT}}{\longrightarrow} \mathbf{Y} \overset{\textcolor{red}{\lceil \cdot/\mathbf{Q}\rfloor}}{\longrightarrow}\mathbf{Y'}}^{E_{\text{JPEG}}}\underbrace{ \overset{\cdot \odot \mathbf{Q}}{\longrightarrow} \tilde{\mathbf{Y}}  \overset{\text{DCT}^{-1}}{\longrightarrow} \mathbf{\tilde X}}_{D_{\text{JPEG}}} ,
\end{align}
where $\mathbf{Y}$ and $\mathbf{Y'}$ indicate discrete cosine transform (DCT \cite{ahmed1974discrete}) spectra and symbols, respectively. We notate $\tilde{(\cdot)}$ as a distorted signals and $\hat{(\cdot)}$ as predictions.
$\mathbf{X}$ is divided into luminance ($\mathbf{X}_Y$) and chroma ($\mathbf{X}_C$) components from RGB, and $\downarrow_{\times 2}$ indicates chroma subsampling. The range resolution of the $\downarrow_{\times 2}$ operation is optional between 4:4:4 ($H \times W$), 4:2:2 ($H \times W/2$), and 4:2:0 ($H/2 \times W/2$). The $E_\text{JPEG}$ quantizes the $8\times 8$ DCT spectrum ($\mathbf{Y}$) with the predefined quantization matrix ($\mathbf{Q}\in [1,255]^{8\times8}$). The $D_\text{JPEG}$ decodes $\mathbf{Y'}$ in reverse order of $E_\text{JPEG}$.  In summary, all JPEG losses originate from quantization and chroma subsampling, highlighted in red in \cref{eq:jpeg_oneline}. 
However, due to the principle of the data processing inequality \cite{informationtheorybook} and the nature of the JPEG, the mutual information $I(\mathbf{X};\mathbf{\tilde X})$ is limited to the $E_\text{JPEG}$.

\noindent\textbf{Neural Decoder} To address this, the Neural JPEG decoder ($:=D_\theta$) learns an image prior from datasets, thereby embedding the image prior into a trainable parameter $\theta$. From the perspective of mutual information, it follows:
\begin{align}
    \label{eq:dec_jpeg}
    &I\bigl(\mathbf{X};D_{\hat\theta}(E_{\text{JPEG}}(\mathbf{X}))\bigr) \sim I(\mathbf{X};\mathbf{\tilde X}) + I(\mathbf{\tilde X};\hat\theta),
\end{align}
\vspace{-20pt}
\small{
\begin{equation}
    \hat{\theta} := \arg\max_{\theta}  \left[\sum_{x\in \mathbf{X},\tilde y\in\mathbf{\tilde Y}} \log p\Bigl(x \mid D_\theta(\tilde y)\Bigr) + \log p(\theta)\right].
    \label{eq:map_dec}
\end{equation}}
\normalsize
\cref{eq:dec_jpeg} indicate that it is equivalent to obtaining additional mutual information from the trained parameters by \cref{eq:map_dec}. 

\noindent\textbf{Neural Encoder} In JPEG encoder, an entropy ($:=\mathcal{H}(\cdot)$) of the spectrum satisfies $\mathcal{H}(\mathbf{Y})\geq \mathcal{H}(\mathbf{ Y'})$, enabling compression. However, the mutual information between an original image and symbols satisfies $I(\mathbf{X};\mathbf{X'})\geq I(\mathbf{X};\mathbf{\tilde X})$, resulting in a loss for the image. We optimize two objectives: 1) increasing $I(\mathbf{X};E_\varphi(\mathbf{X}))$ and 2) minimizing $\mathcal{H}(\mathbf{Y'})$ with the Neural JPEG encoder ($:=E_\varphi$).
\vspace{-10pt}
\begin{align}
\label{eq:enc_jpeg}
    I(\mathbf{X};D_\text{JPEG}(E_\varphi(\mathbf{X}))) &\geq I(\mathbf{X};\mathbf{\tilde{X}}), \\
    \text{s.t. } \mathcal{H}(\mathbf{X})\geq  \mathcal{H}({E_\text{JPEG}(\mathbf{X})})&\geq  \mathcal{H}(E_\varphi(\mathbf{X})).
\end{align}
To satisfy the \cref{eq:enc_jpeg}, we add mutual information in a manner of \cref{eq:dec_jpeg} such as:
\begin{align}
&I\bigl(\mathbf{X};E_\varphi(\mathbf{X})\bigr) \sim I(\mathbf{X};\mathbf{X'}) + I(\mathbf{X'};\varphi),
    \label{eq:enc_addd_info}
\end{align}
\vspace{-20pt}
\begin{align}
    &\hat{\varphi} := \arg\max_{\varphi}  \left[\sum_{x\in \mathbf{X}} \log p\Bigl(x \mid  x'\Bigr) + \log p(\varphi)\right].
    \label{eq:map_enc}
\end{align}
In \cref{eq:enc_addd_info}, we add mutual information from the trained parameters by \cref{eq:map_enc}. 
In other words, our method focuses on specifically addressing the errors introduced during chroma subsampling (i.e. $\downarrow_{\times 2}$). The continuous function property of INRs, which benefits image upsampling and subsampling \cite{pan2022towards}, is extended to the neural operator while preserving these advantages.

\section{Method}\label{method}
\begin{figure}[t]
\footnotesize
\centering
\includegraphics[trim= 0 0 0 0,clip,width = 3.3in]{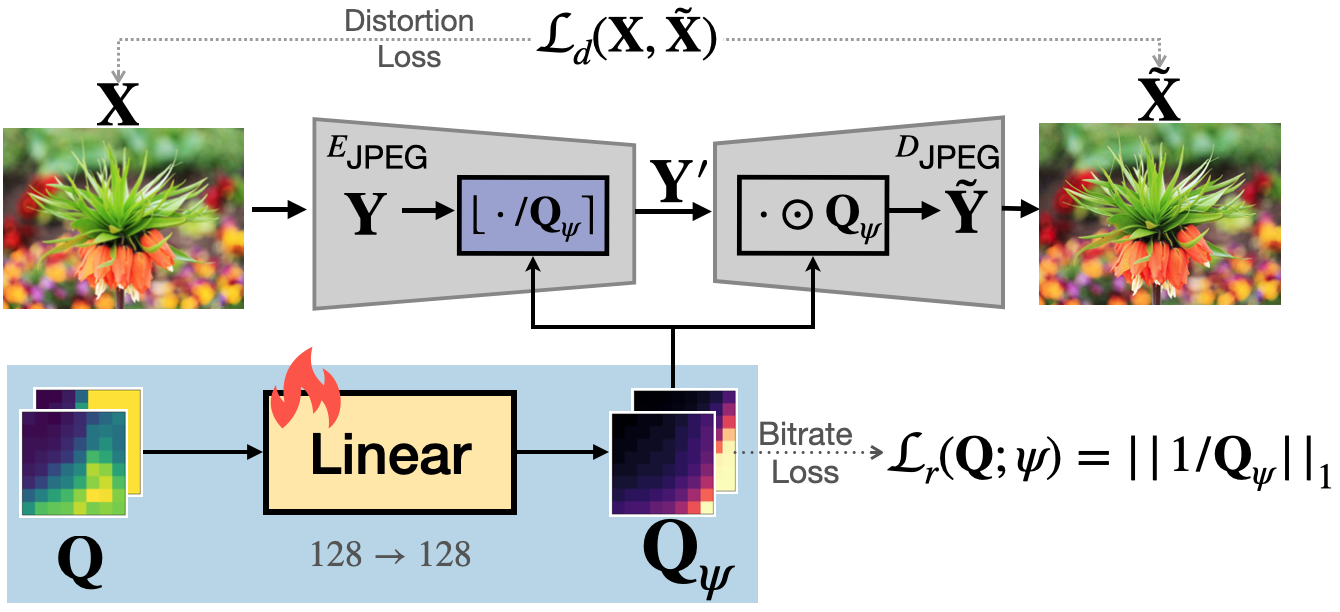}
\vspace{-17pt}
\caption{\textbf{Overview of learning the quantization matrix.} Only one linear layer is trained. Since quantization is not differentiable, it is replaced with an approximation operation (blue area).}
\label{fig:qmap_learning}
\vspace{-20pt}
\end{figure}

\noindent\textbf{Training Quantization Matrix ($\mathbf{Q}_\psi$)} Inspired from \citet{learnablejpeg}, our JPNeO uses the pre-trained quantization matrix instead of using the existing quantization matrix. \cref{fig:qmap_learning} illustrates training procedure of $\mathbf{Q}_\psi$. 
The round operation ($\lfloor \cdot \rceil$) banishes gradients of the network. Therefore, it is replaced to 3rd order approximation:
\begin{align}
    \lfloor x \rceil \simeq \lfloor x \rceil + (\lfloor x \rceil-x)^3.
\end{align}
Unlike existing work \cite{learnablejpeg} that uses trainable parameters, we use a standard quantization map (quality factor = 50) as an input of the linear layer which is equivalent to initializing the parameters (i.e. $\mathbf{Q}_\psi =\text{Linear}(\mathbf{Q})\ ; \text{Linear}:\mathbb{R}^{128}\mapsto\mathbb{R}^{128}$).
The loss function is calculated using the following formula:
\begin{align}
        \mathcal{L}(\mathbf{X},\mathbf{\tilde{X}}|\mathbf{Q}_\psi) &= \lambda \cdot \mathcal{L}_d(\mathbf{X},\mathbf{\tilde{X}}) + \mathcal{L}_r(\mathbf{Q};\psi),\\
 \mathcal{L}_d(\mathbf{X},\mathbf{\tilde{X}}):=& ||\mathbf{X}-\mathbf{\tilde{X}}||_2,\  \mathcal{L}_r(\mathbf{Q};\psi) := ||1/\mathbf{Q}_\psi||_1.
    \label{eq:loss_qmap}
\end{align}
The hyperparameter $\lambda$ controls the ratio of the distortion loss function ($\mathcal{L}_d(\cdot,\cdot)$), and the bitrate loss function ($ \mathcal{L}_r(\cdot)$) for the determination of the compression ratio.
One model is trained per $\lambda$. Given that $\mathbf{Q}_\psi$ is not large, we found that it is effective to store the result $\mathbf{Q}_\psi$ after training and use them as a lookup table.

\begin{figure}[t]
\footnotesize
\centering
\includegraphics[trim= 0 0 0 0,clip,width = 3.3in]{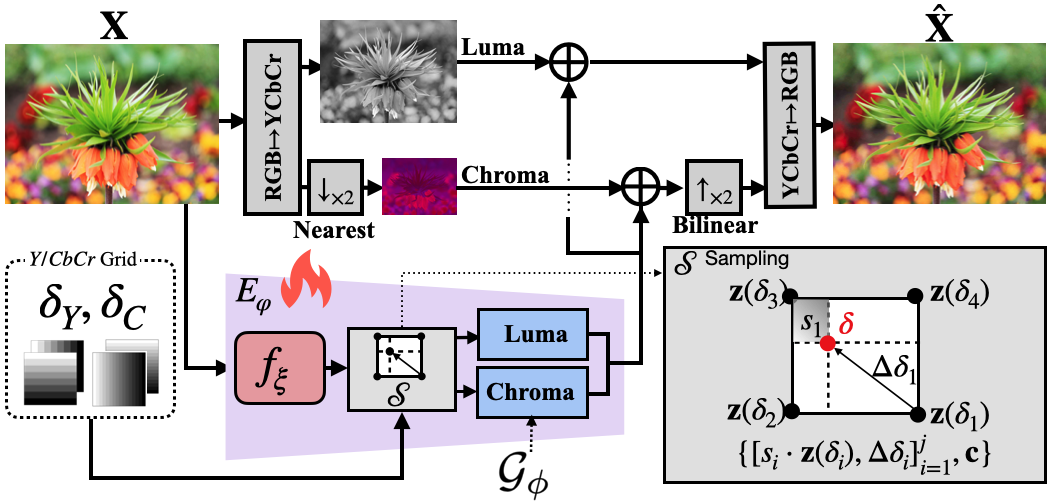}
\vspace{-17pt}
\caption{\textbf{Schematic flow of training our JENO.} JENO takes $\mathbf{X}$ and coordinates of the $\mathbf{X}$ and downsampled chroma image as input. The predicted image of JENO is produced by a JPEG decoder without the quantization process.}
\label{fig:train_jeno}
\vspace{-10pt}
\end{figure}

\noindent\textbf{JPEG Encoding Neural Operator (JENO)} We propose a novel JPEG encoding neural operator that addresses chroma subsampling. The training procedure of our JENO is illustrated in \cref{fig:train_jeno}.
 Our JENO ($E_\varphi: (\mathbf{X},(\delta_Y,\delta_C))\mapsto (\mathbf{X}_Y,\mathbf{X}_C)$) takes RGB image ($\mathbf{X}$) and coordinate of Y/CbCr components ($:=\delta_{Y,C}\in{ \mathbb{R}^{r_1H\times r_2W\times2}}$) as inputs. Note that $r_{1,2}\in\{0.5,1\}$ are determined by the subsampling mode of JPEG. JENO consists of the feature extractor ($f_\xi:\mathbf{X} \mapsto \mathbf{z}\in \mathbb{R}^{H\times W \times K}$), sampling \cite{chen2021learning,lee2021local} ($\mathcal{S}: (\mathbf{z},\delta)\mapsto \mathbf{z}_0\in\mathbb{R}^{H'\times W'\times K'}$) and Galerkin-Attention module ($G_\phi$) \cite{wei2023super, kovachki2023neural}.
Our JENO utilizes a neural operator as an auxiliary encoder for JPEG such that:
\begin{align}
    \mathbf{\hat X} = \mathcal{G}_\phi(\mathcal{S}(f_\xi(\mathbf{X}),\delta)) + \mathbf{X'}, 
    \label{eq:jeno_main}
\end{align}
\vspace{-20pt}
\begin{align}
    \mathcal{S}(\mathbf{z},\delta) := \{ [s_i\cdot\mathbf{z}(\delta_i),\Delta\delta_i]^{j}_{i=1},\mathbf{c}\},
\end{align}
\vspace{-20pt}

\noindent where $\delta_i$ are the coordinates of the nearest for neighbors of $\delta$, $\Delta\delta = \delta-\delta_i$, $\mathbf{c} = (2/r_1,2/r_2)$, and $s_i = \Delta\delta_{i,\mathbf{h}}\times \Delta\delta_{i,\mathbf{w}}$ local area of an image.
We optimize a set of trainable parameters $\varphi=\{\xi;\phi\}$ as below:
\begin{align}
    \hat{\varphi} := \arg\min_\varphi ||\mathbf{X} - U(\mathbf{\hat X_\varphi)} ||_1,
\end{align}
where $U(\cdot)$ indicates a bilinear upsampling. As a result, our JENO learns high-frequency components of $\mathbf{X}$ by rewriting \cref{eq:jeno_main} as below:
\begingroup
\small
\begin{align}
    \mathbf{\hat X} &= U(E_\varphi(\mathbf{X})) + U(\mathbf{X}')\\
    \ &\simeq HPF(\mathbf{X}) + LPF(\mathbf{X}) \\
    \therefore U(E_\varphi(\mathbf{X})) &\simeq HPF(\mathbf{X}) \quad (\because U(\mathbf{X}') =LPF(\mathbf{X})),
    \label{eq:hpf}
\end{align}
\endgroup
\vspace{-20pt}

\noindent where $HPF, LPF$ are a high-pass filter and a low-pass filter, respectively. 

\begin{figure}[!t]
    \centering
    \includegraphics[width=3.2in]{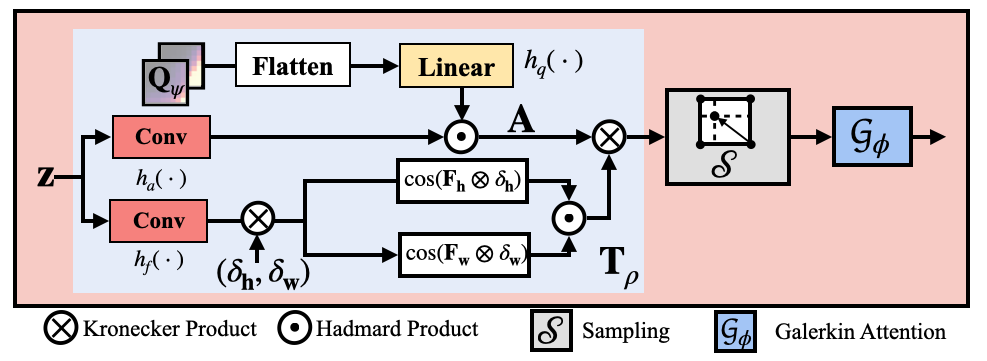}
        \vspace{-10pt}
    \caption{\textbf{Schematic flow of Cosine Neural Operator (CNO).}}
    \label{fig:CCF-detail}
    \vspace{-10pt}
\end{figure}

\begin{figure*}[!t]
\centering
\scriptsize
    \stackunder[2pt]{}{\includegraphics[trim=0 0 0 0,clip,width=0.95\linewidth]{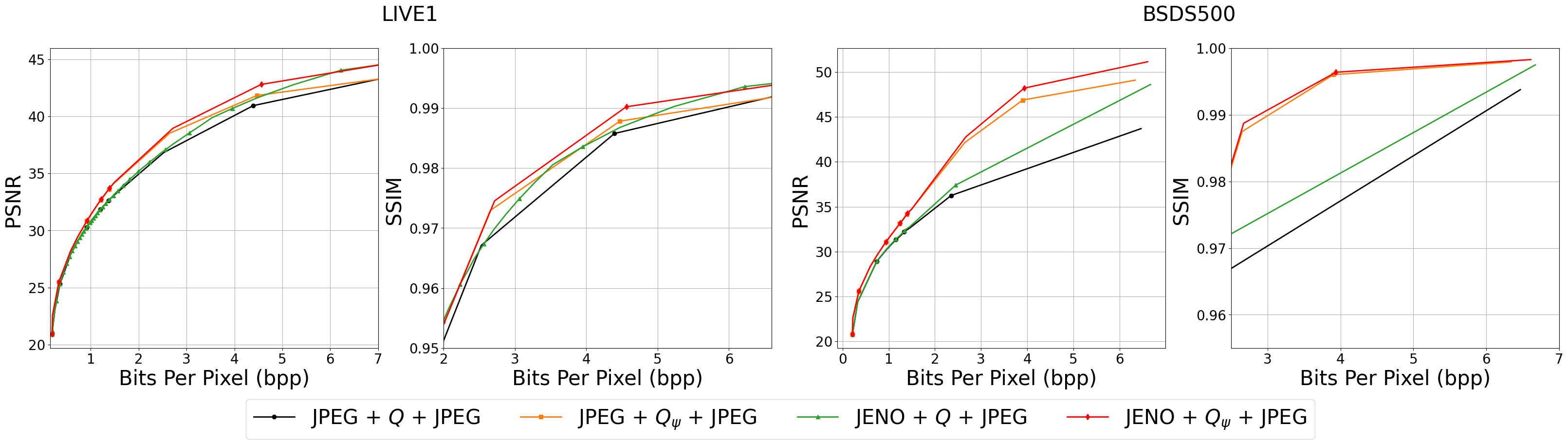}}
     \stackunder[2pt]{}{\includegraphics[trim=0 0 0 0,clip,width=0.95\linewidth]{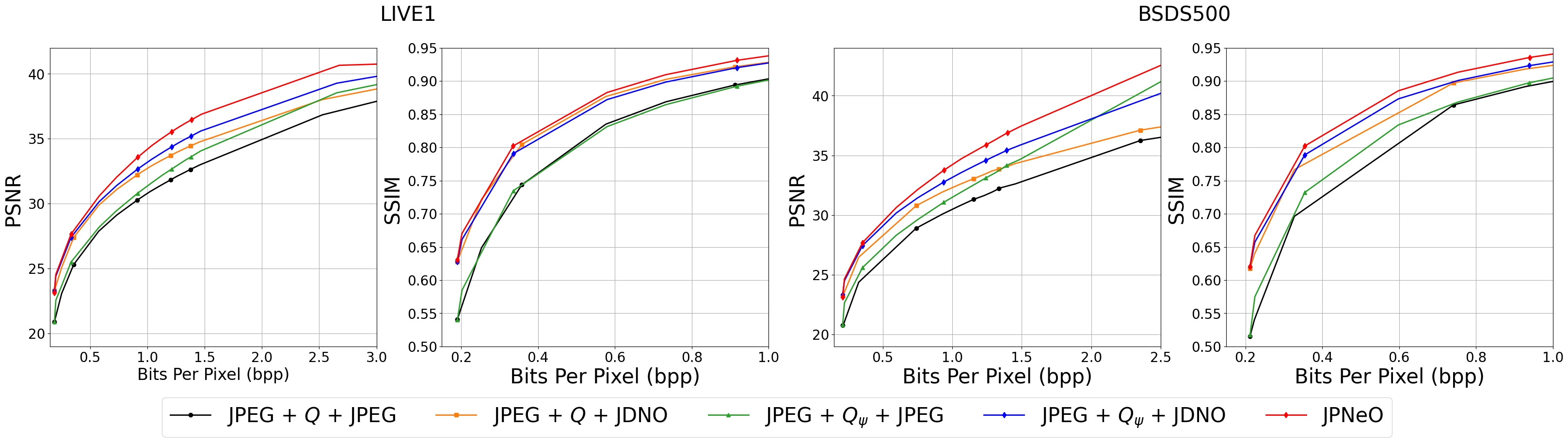}}
    \vspace{-10pt}
    \caption{Quantitative ablation study on our JENO (top), JDNO (bottom), and $\mathbf{Q}_\psi$ (PSNR($\uparrow$)/SSIM($\uparrow$)). The label indicates `Encoder ($E_\varphi/E_\text{JPEG}$)'$+$`$\mathbf{Q/Q}_\psi$'$+$Decoder ($D_\theta/D_\text{JPEG}$)'. The RD curves were measured on validation set of LIVE1 \cite{Live1} and BSDS500 \cite{B500dataset} dataset. The 4:2:0 chroma downsampling was used.}
    \vspace{-20pt}
    \label{fig:rd_self}
\end{figure*}
\noindent\textbf{JPEG Decoding Neural Operator (JDNO)} For decoding symbols, we introduce JDNO ($D_\theta:(\mathbf{Y}_Y,\mathbf{Y'}_C;\mathbf{Q})\mapsto \mathbf{\hat X}$).
Our JDNO is inspired by the JPEG decoding INR proposed by \citet{jdec2024han}; however, JDEC \cite{jdec2024han} does not support 4:2:2 or 4:4:4 chroma subsampling. JDNO consists of three parts: 1) the group embedding ($g:(\mathbf{\tilde Y}_Y,\mathbf{\tilde Y}_C)\mapsto \mathbf{z}\in\mathbb{R}^{H/B\times W/B \times3 B^2}$), 2) the feature extractor ($f_\xi$), and 3) the neural operator with continuous cosine function $(\mathcal{G_\psi \circ}T_\rho)$. 
$g$ takes chroma and luma spectra as input and embeds them as below:
\small{
\begin{equation}
g(\mathbf{\tilde Y}_Y,\mathbf{\tilde Y}_C)\ := [\mathcal{D}_B(\mathcal{D}^{-1}_8(\tilde{\mathbf{Y}}_Y)), \mathcal{D}_B(\mathcal{D}^{-1}_8(U' \cdot \tilde{\mathbf{Y}}_C))] 
\end{equation}}
\normalsize
where $\mathcal{D}_N$ indicates 2D-DCT with size of $N\times N$ and $U'$ is an impulse response of $U$. In $f_\xi:\mathbf{z} \mapsto \mathbf{z}'\in \mathbb{R}^{H/B\times W/B \times K}$, feature is extracted as in JENO. We utilize continuous cosine formulation \cite{jdec2024han} into Cosine Neural Operator (CNO) which comprises three modules: $h_f \colon \mathbb{R}^K\mapsto \mathbb{R}^{2M}$, coefficient estimator $h_a \colon \mathbb{R}^K\mapsto \mathbb{R}^M$, and quantization matrix encoder $h_q \colon \mathbb{R}^{128} \mapsto \mathbb{R}^M$. 
Then, CNO formulate the latent vector $\mathbf{z}'$ as below:
\vspace{-5pt}
\begingroup
\small
\begin{align}
 \mathbf{T}_\rho(\mathbf{z}',\delta;\mathbf{Q}) =& \mathbf{A\otimes(\cos(\pi\mathbf{F}_{h}\otimes\delta_h)\odot \cos(\pi \mathbf{F}_{w}\otimes\delta_w))} \\
    \text{where }& \mathbf{A} =h_q(\mathbf{Q}) \odot h_a(\mathbf{z}'),\ \mathbf{F} =h_f(\mathbf{z}').
\end{align} 
\endgroup
 $\odot\ , \otimes$ denote Hadamard and Kronecker product, respectively. \cref{fig:CCF-detail} shows overall flow of our CNO.
$\mathbf{\hat X}$ is completed by decoding $\mathbf{z}_0 = \mathbf{T}\rho(\mathbf{z}',\delta;\mathbf{Q})$ JENO's neural operator process (\cref{eq:jeno_main}) i.e:
\begin{align}
    \mathbf{\hat X} = \mathcal{G}_\phi(\mathcal{S}\circ\mathbf{T}_\rho(f_\xi(\mathbf{X}),\delta)).
\end{align}
The trainable parameters of JDNO $\mathbf{\Theta} := \{\xi ;\rho;\phi \}$ is optimized with the equation below:
\begin{equation}
      \widehat\Theta = \arg\min_{\Theta} || \mathbf{X} - \hat{\mathbf{X}}_\Theta ||_{1}.
  \end{equation}

\begin{figure*}[t]
\footnotesize
\centering
    \renewcommand{\arraystretch}{1.2}
    \setlength{\tabcolsep}{1pt} 
    \begin{tabular}{ccccccc}
        \stackunder[2pt]{\stackunder[3pt]{JPEG+$\mathbf{Q}$+JPEG}{\includegraphics[trim=90 0 90 60,clip,width=1.09in]{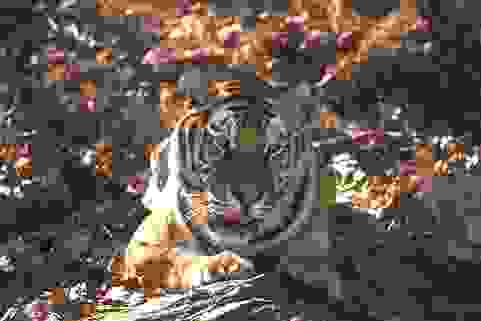}}}{0.2618/19.91/0.559} &
        \stackunder[2pt]{\stackunder[2pt]{JPEG+$\mathbf{Q}_\psi$+JPEG}{\includegraphics[trim=90 0 90 60,clip,width=1.09in]{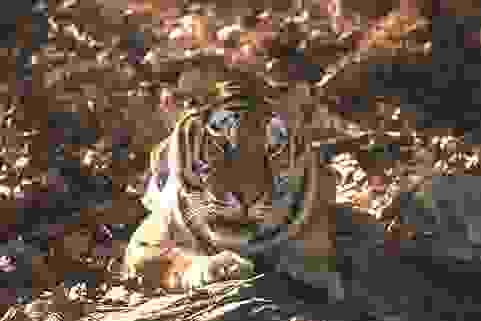}}}{\textcolor{blue}{0.2602}/21.21/0.581} &
        \stackunder[2pt]{\stackunder[3pt]{JPEG+$\mathbf{Q}$+JDNO}{\includegraphics[trim=90 0 90 60,clip,width=1.09in]{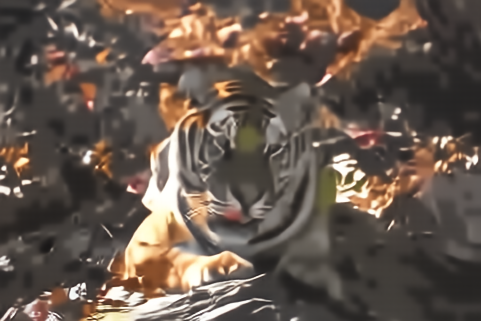}}}{0.2618/\textcolor{black}{22.27}/\textcolor{black}{0.643}} &
        \stackunder[2pt]{\stackunder[2pt]{JPEG+$\mathbf{Q}_\psi$+JDNO}{\includegraphics[trim=90 0 90 60,clip,width=1.09in]{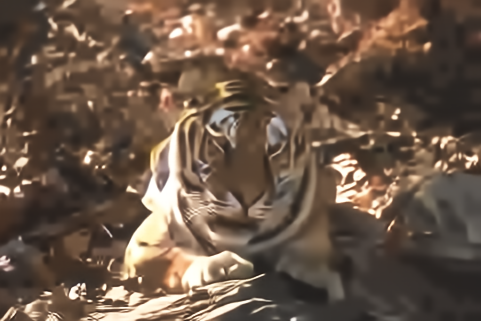}}}{\textcolor{blue}{0.2602}/\textcolor{blue}{23.10}/\textcolor{blue}{0.661}} &
        \stackunder[2pt]{\stackunder[4pt]{\textbf{JPNeO}}{\includegraphics[trim=90 0 90 60,clip,width=1.09in]{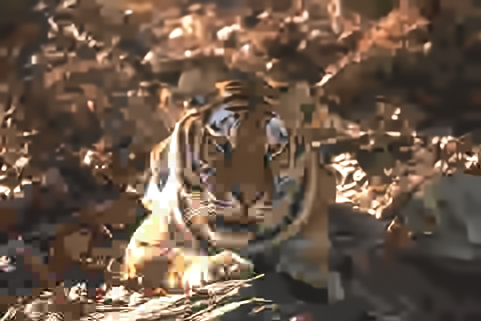}}}{\textcolor{red}{0.2597}/\textcolor{red}{23.36}/\textcolor{red}{0.680}} &
        \stackunder[2pt]{\stackunder[4pt]{Ground Truth}{\includegraphics[trim=90 0 90 60,clip,width=1.09in]{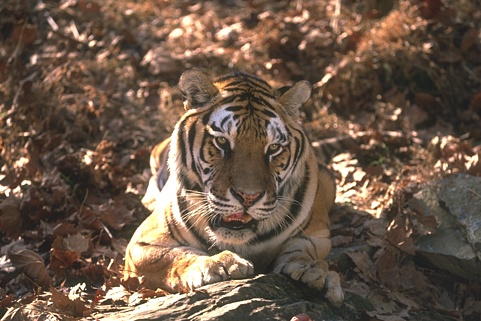}}}{bpp($\downarrow$)/PSNR($\uparrow$)/SSIM($\uparrow$)} \\
        
        \stackunder[3pt]{\includegraphics[trim=300 50 90 200,clip,width=1.088in]{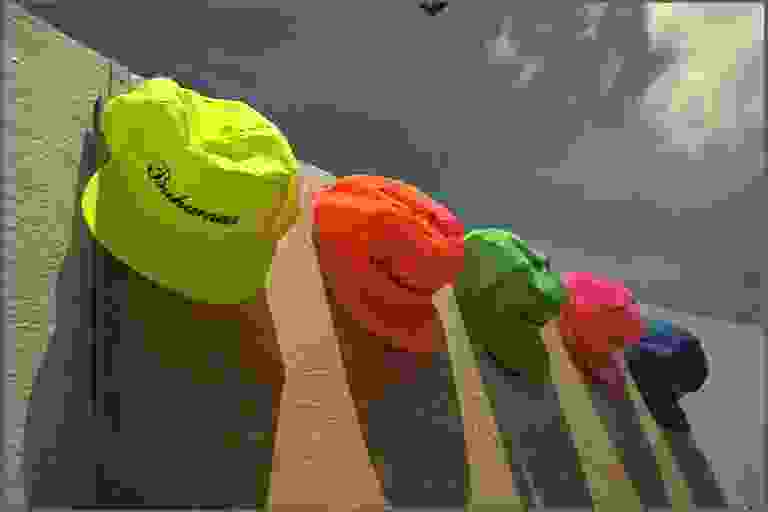}}{0.1668/23.71/0.667} &
        \stackunder[3pt]{\includegraphics[trim=300 50 90 200,clip,width=1.088in]{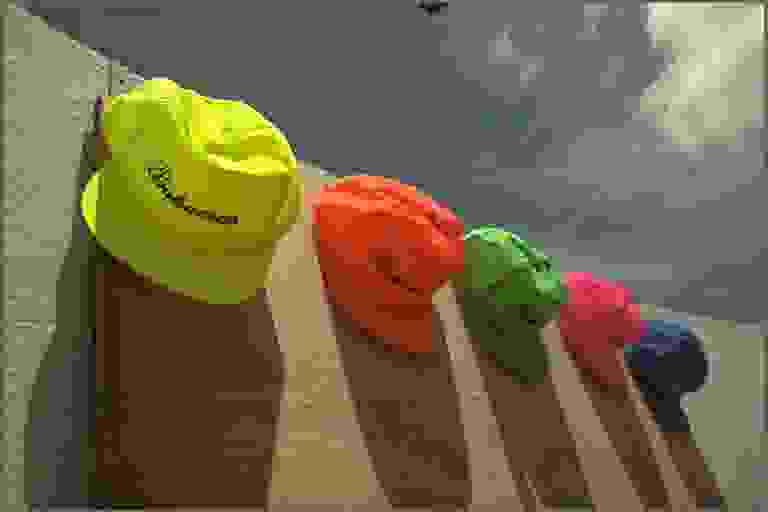}}{\textcolor{blue}{0.1606}/25.31/0.676} &
        \stackunder[3pt]{\includegraphics[trim=300 50 90 200,clip,width=1.088in]{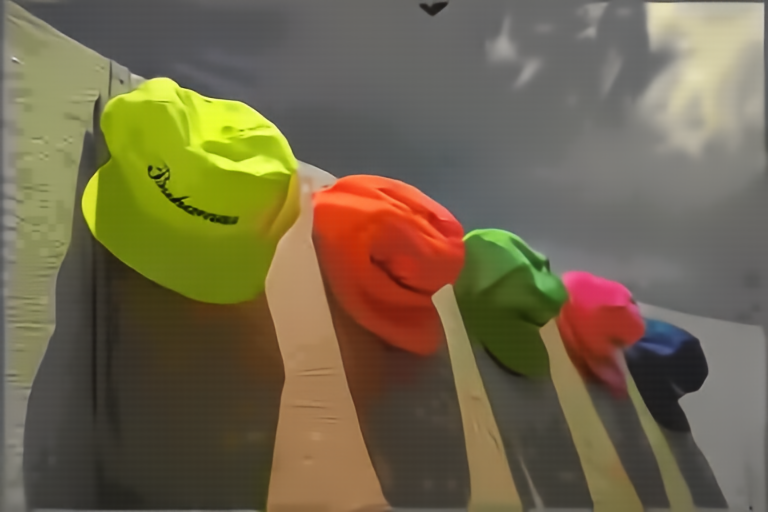}}{0.1668/\textcolor{black}{27.17}/\textcolor{black}{0.788}} &
        \stackunder[3pt]{\includegraphics[trim=300 50 90 200,clip,width=1.088in]{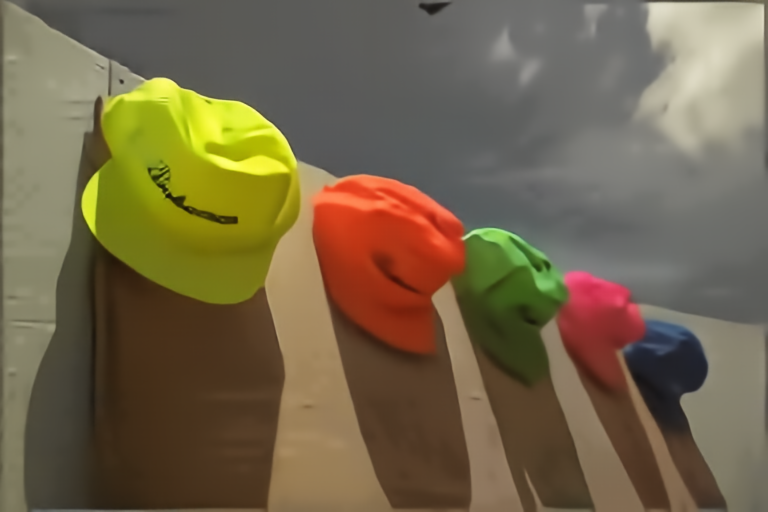}}{\textcolor{blue}{0.1606}/\textcolor{blue}{28.06}/\textcolor{blue}{0.789}} &
        \stackunder[3pt]{\includegraphics[trim=300 50 90 200,clip,width=1.088in]{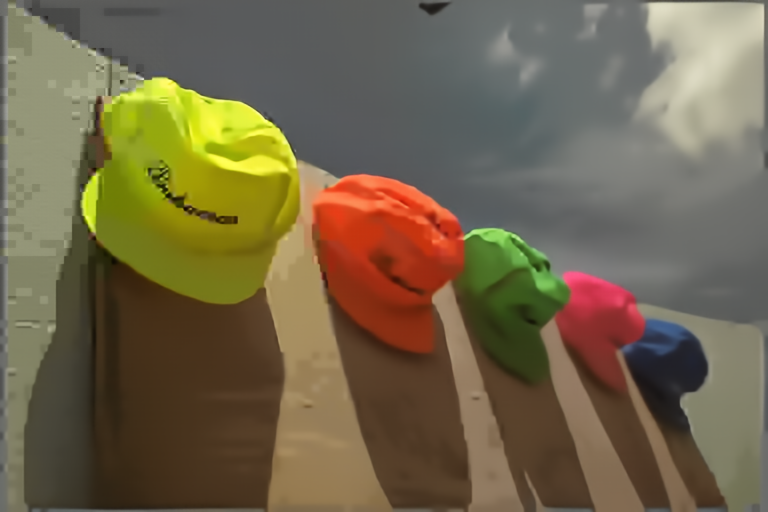}}{\textcolor{red}{0.1605}/\textcolor{red}{28.45}/\textcolor{red}{0.790}} &
        \stackunder[2pt]{\includegraphics[trim=300 50 90 200,clip,width=1.088in]{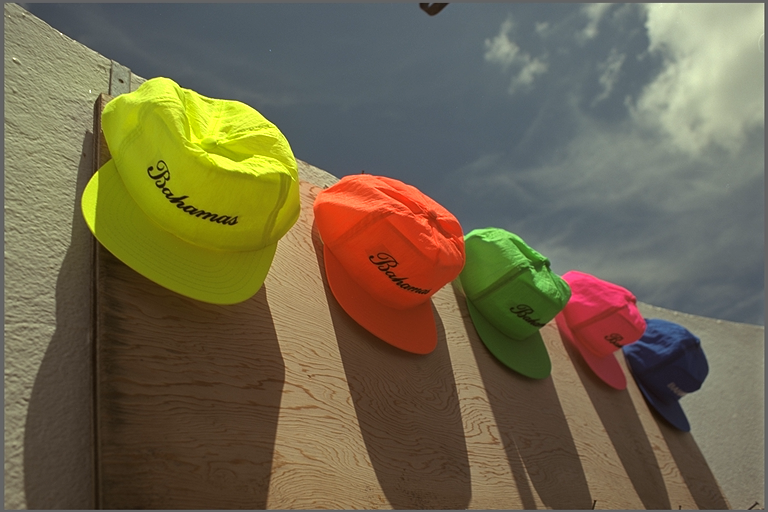}}{bpp($\downarrow$)/PSNR($\uparrow$)/SSIM($\uparrow$)}
    \end{tabular}
    
    \vspace*{-6pt}
    \caption{Qualitative ablation study in highly compressed images based on the quantization map ($\mathbf{Q}/\mathbf{Q}_\psi$) and decoder (JPEG/JDNO). \textcolor{red}{Red} and \textcolor{blue}{blue} colors indicate the best and the second-best performance, respectively.}
    \vspace*{-10pt}
    \label{fig:Qual_self_jpeg}
\end{figure*}

\begin{figure*}[t]
\footnotesize
\centering

\stackunder[2pt]{\stackunder[2pt]{JPEG+$\mathbf{Q}$+JPEG}{\begin{tikzpicture}[x=6cm, y=6cm, spy using outlines={every spy on node/.append style={smallwindow_w}}]
\node[anchor=south] (FigA) at (0,0)  {\includegraphics[trim=160 226 508 189,clip,width=0.135\linewidth]{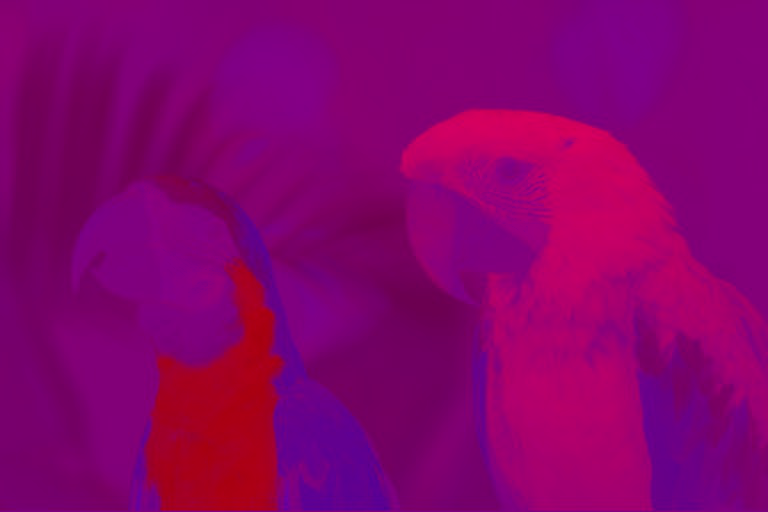}};
\end{tikzpicture}}}{1.465/39.30/0.955}
\hspace{-7pt}
\stackunder[2pt]{\stackunder[1pt]{JPEG+$\mathbf{Q}_\psi$+JPEG}{\begin{tikzpicture}[x=6cm, y=6cm, spy using outlines={every spy on node/.append style={smallwindow_w}}]
\node[anchor=south] (FigA) at (0,0)  {\includegraphics[trim=160 226 508 189,clip,width=0.135\linewidth]{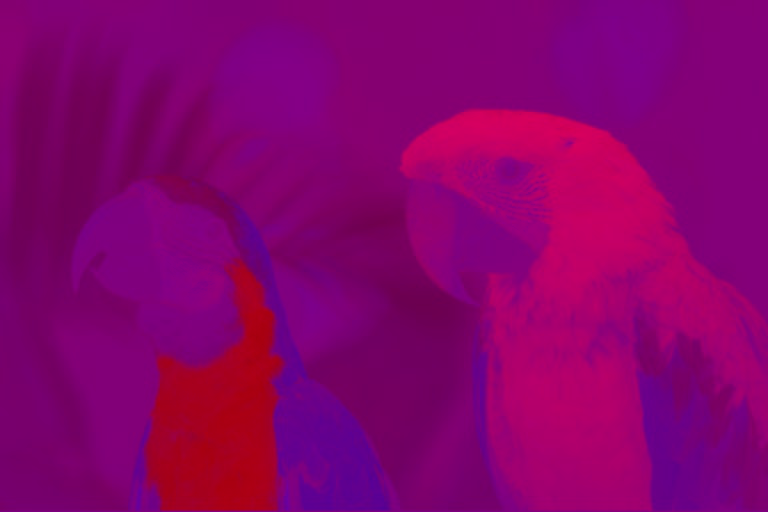}};
\end{tikzpicture}}}{\textcolor{blue}{1.387}/\textcolor{black}{40.04}/0.958}
\hspace{-7pt}
\stackunder[2pt]{\stackunder[2pt]{JENO+$\mathbf{Q}$+JPEG}{\begin{tikzpicture}[x=6cm, y=6cm, spy using outlines={every spy on node/.append style={smallwindow_w}}]
\node[anchor=south] (FigA) at (0,0)  {\includegraphics[trim=160 226 508 189,clip,width=0.135\linewidth]{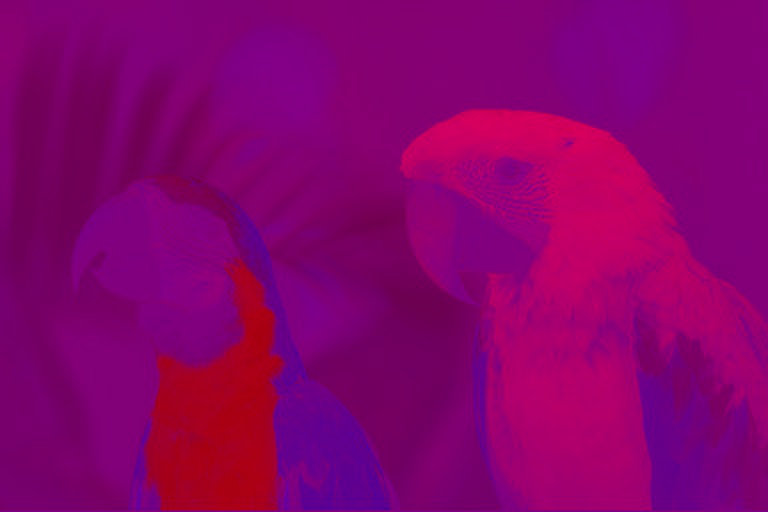}};
\end{tikzpicture}}}{1.462/39.60/\textcolor{black}{0.964}}
\hspace{-7pt}
\stackunder[2pt]{\stackunder[1pt]{JENO+$\mathbf{Q}_\psi$+JPEG}{\begin{tikzpicture}[x=6cm, y=6cm, spy using outlines={every spy on node/.append style={smallwindow_w}}]
\node[anchor=south] (FigA) at (0,0)  {\includegraphics[trim=160 226 508 189,clip,width=0.135\linewidth]{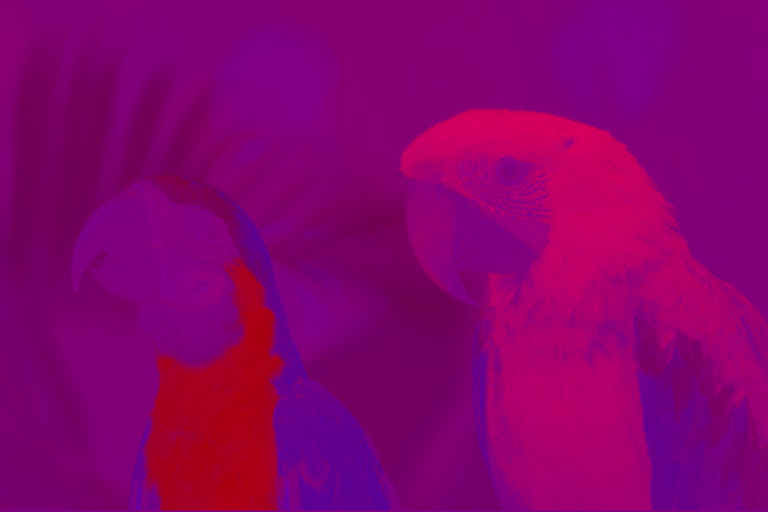}};
\end{tikzpicture}}}{\textcolor{black}{1.408}/\textcolor{blue}{40.23}/\textcolor{blue}{0.965}}
\hspace{-7pt}
\stackunder[2pt]{\stackunder[3pt]{\textbf{JPNeO}}{\begin{tikzpicture}[x=6cm, y=6cm, spy using outlines={every spy on node/.append style={smallwindow_w}}]
\node[anchor=south] (FigA) at (0,0)  {\includegraphics[trim=160 226 508 189,clip,width=0.135\linewidth]{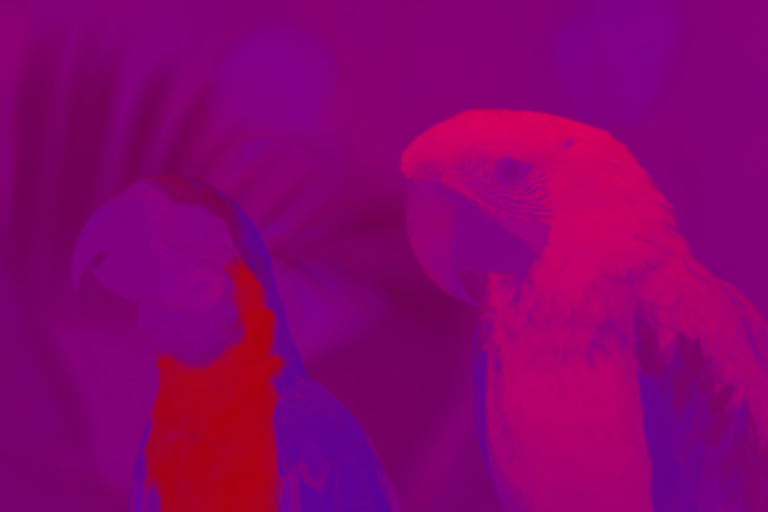}};
\end{tikzpicture}}}{\textcolor{red}{1.381}/\textcolor{red}{43.02}/\textcolor{red}{0.977}}
\hspace{-7pt}
\stackunder[2pt]{\stackunder[3.5pt]{Ground Truth}{\begin{tikzpicture}[x=6cm, y=6cm, spy using outlines={every spy on node/.append style={smallwindow_w}}]
\node[anchor=south] (FigA) at (0,0)  {\includegraphics[trim=160 226 508 189,clip,width=0.135\linewidth]{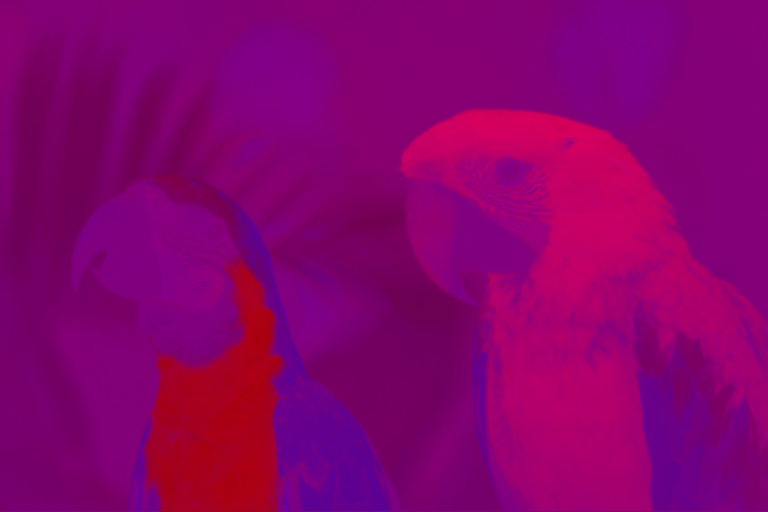}};
\end{tikzpicture}}}{Chroma}
\hspace{-7pt}
\stackunder[2pt]{\stackunder[3.5pt]{Ground Truth}{\begin{tikzpicture}[x=6cm, y=6cm, spy using outlines={every spy on node/.append style={smallwindow_w}}]
\node[anchor=south] (FigA) at (0,0)  {\includegraphics[trim=160 226 508 189,clip,width=0.135\linewidth]{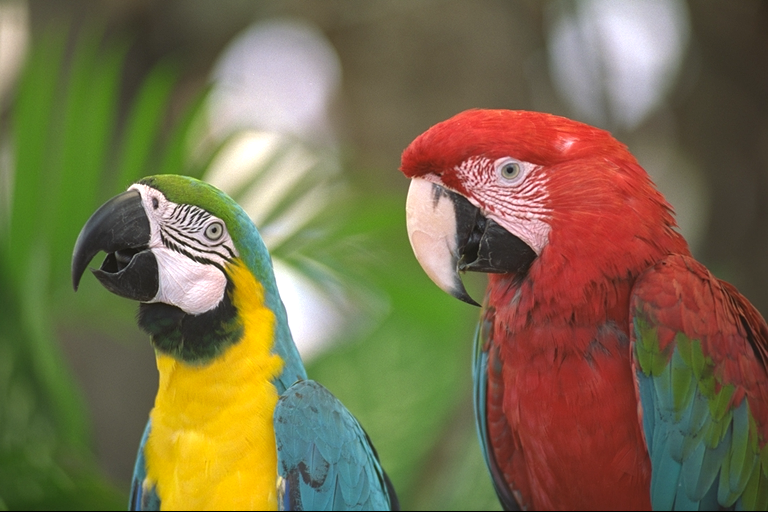}};
\end{tikzpicture}}}{\scriptsize{bpp($\downarrow$)/PSNR($\uparrow$)/SSIM($\uparrow$)}}

\vspace*{-6pt}
\caption{Qualitative comparison of chroma components in high-bpp images based on the encoder (JENO) and the quantization matrix ($\mathbf{Q}$/$\mathbf{Q}_\psi$). \textcolor{red}{Red} and \textcolor{blue}{blue} colors indicate the best and the second-best performance, respectively.}
\vspace*{-12pt}
\label{fig:qual_chroma_abl_self}
\end{figure*}

\section{Experiments}
\subsection{Network Implementation Details}
\noindent \textbf{Quantization matrix ($Q_\psi$)} Only one linear layer is trained for $\mathbf{Q}_\psi$, and it is not used after trained. We apply a scaled sigmoid layer ($:=254\cdot sig(\cdot)+1$) to limit the output value. After training, $Q_\psi$ is stored as an integer [1,255]. 
We set $\lambda$ by sampling the values from [$1e^{-4},1e^6$]. As a result, we save 17 different $\mathbf{Q}_\psi$, excluding duplicate values.

\noindent \textbf{JENO ($E_\varphi$)} We use the EDSR-baseline \cite{Lim_2017_CVPR_Workshops} which has 16 residual blocks with 64 ($=K$) channels for feature extractor ($f_\xi$) of our JENO. Therefore, the latent vector for the neural operator contains 266 ($=K'$) channels. Following the configuration of \cite{wei2023super}, $\mathcal{G}_\phi$ comprises 256 channels with 16 heads, and two iterations for the kernel integral.

\noindent \textbf{JDNO ($D_\theta$)} Following the previous decoding method \cite{jdec2024han}, we adopt the SwinV2 attention module \cite{liang2021swinir} with 256 channels ($=K$) as feature extractor ($f_\xi$). For group-embedding ($g(\cdot,\cdot)$), we set the block size as 4 ($=B$). In JDNO, the cosine features ($T_\rho(\mathbf{z})$) contain 128 channels ($=M$), resulting in a total of 512 input channels for the neural operator ($\mathcal{G}_\phi$).
$h_f$ and $h_a$ consist of a single $3\times3$ convolutional layer, while $h_q$ is a single linear layer. $\mathcal{G}_\phi$ performs two iterations with 16 heads, following the same configuration as in JENO.

\begin{figure*}[!t]
    \centering
    \includegraphics[trim=0 0 0 0,clip,width = 0.95\linewidth]{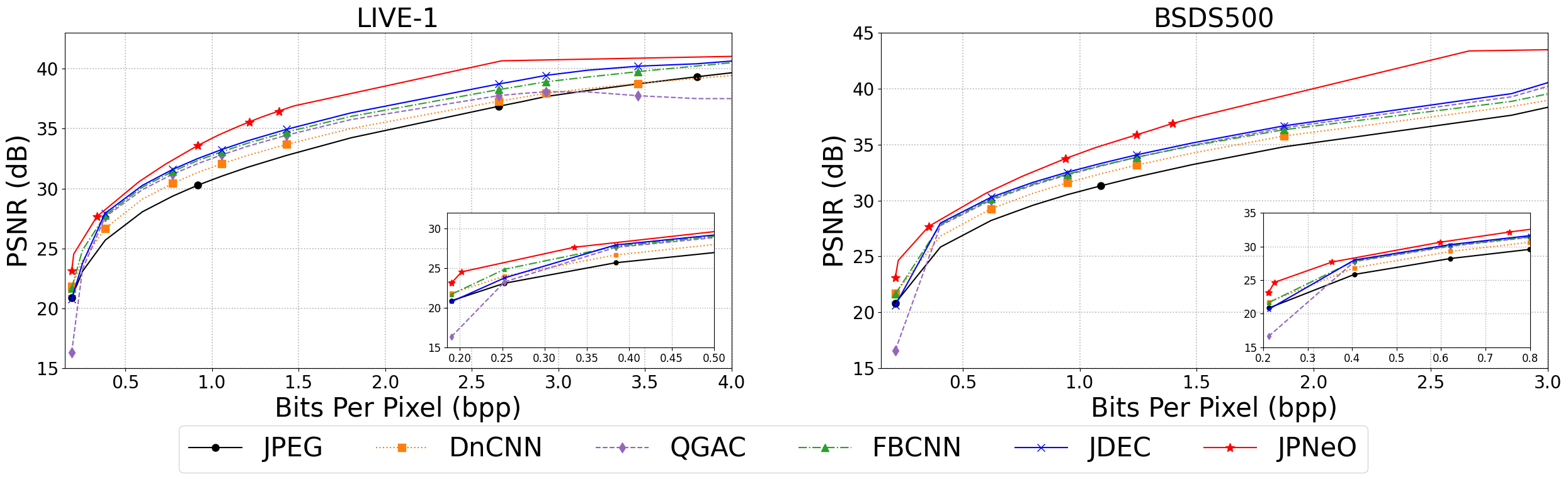}
    \vspace{-10pt}
    \caption{Rate-Distortion cuvre comparisons with the \textit{JPEG artifact removal networks} on LIVE-1 \cite{Live1} (left) and BSDS500 \cite{B500dataset} (right). We highlight the highly compressed parts in the bottom right part of each graph. We show PSNR as a measure of \textit{distortion} (higher is better).}
        \vspace{-15pt}
    \label{fig:RDcurve-qual}
\end{figure*}

\begin{figure*}[t]
\footnotesize
\centering
\hspace{-18pt}

\hspace{-4pt}
\raisebox{0.2in}{\rotatebox{90}{$q=5$}}
{\includegraphics[trim=0 206 276 0,clip,width = 1.09in]{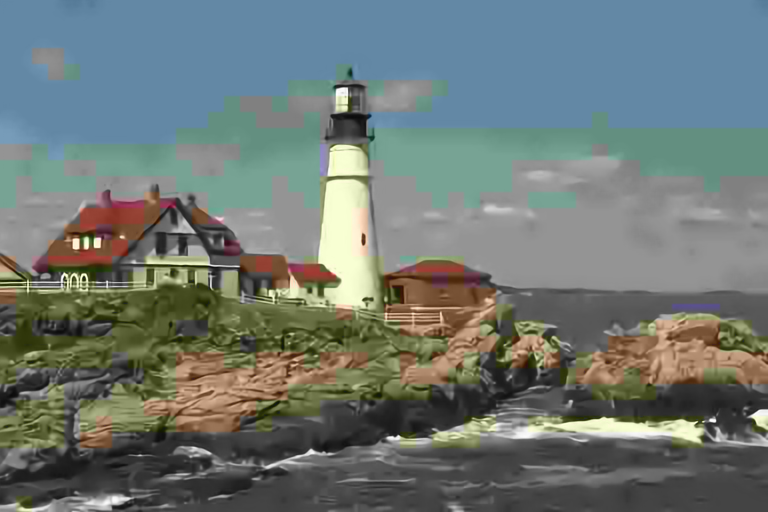}}
{\includegraphics[trim=0 206 276 0,clip,width = 1.09in]{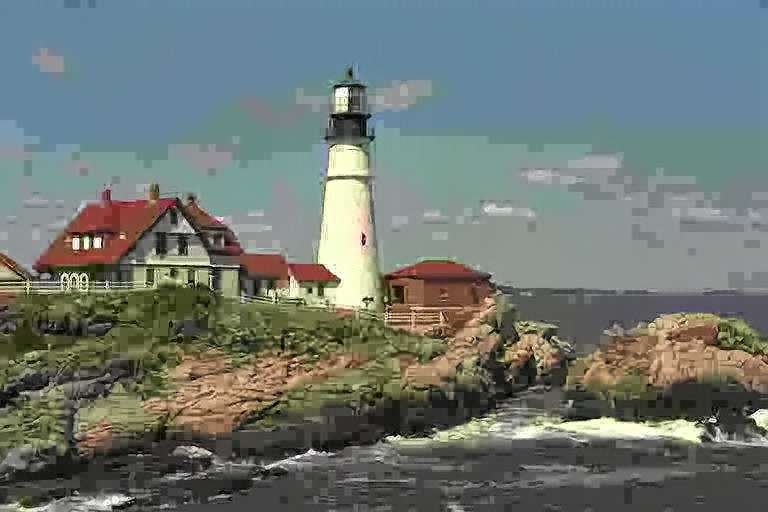}}
{\includegraphics[trim=0 206 276 0,clip,width = 1.09in]{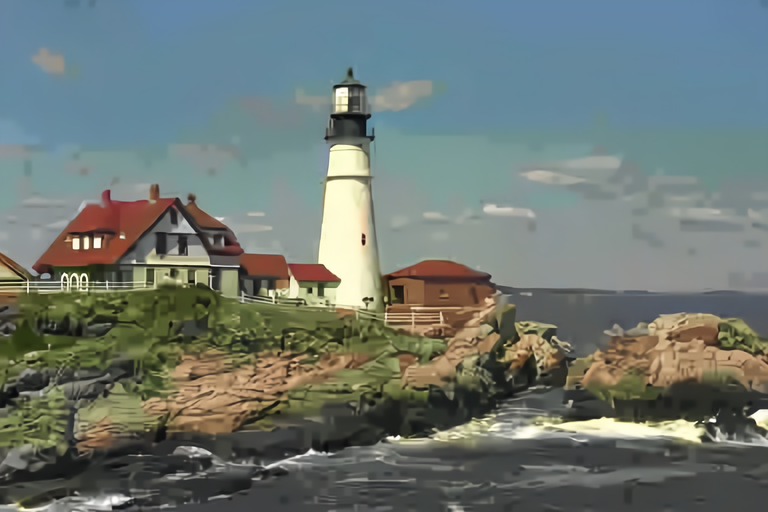}}
{\includegraphics[trim=0 206 276 0,clip,width = 1.09in]{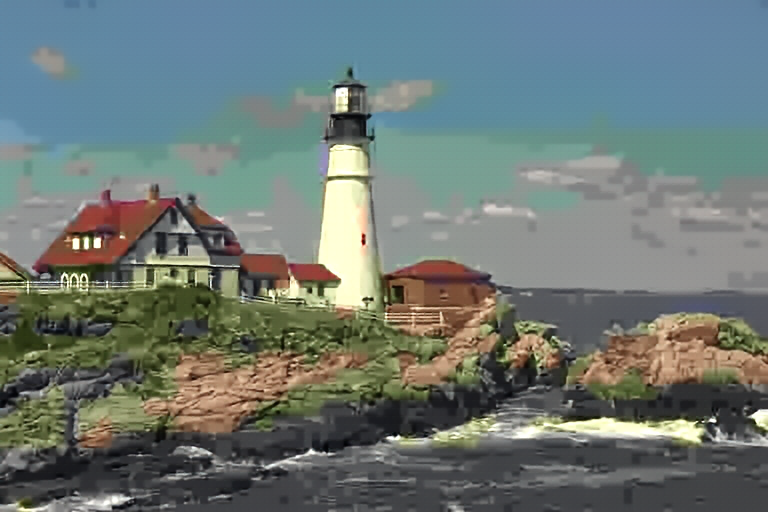}}
{\includegraphics[trim=0 206 276 0,clip,width = 1.09in]{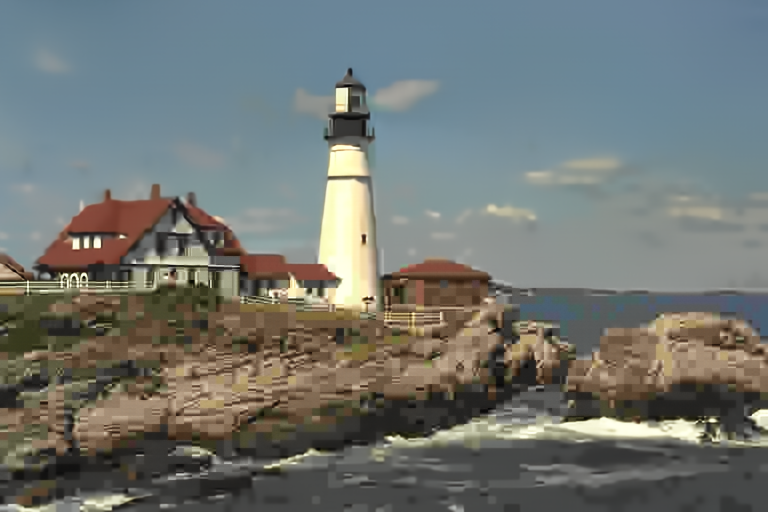}}
{\includegraphics[trim=0 206 276 0,clip,width = 1.09in]{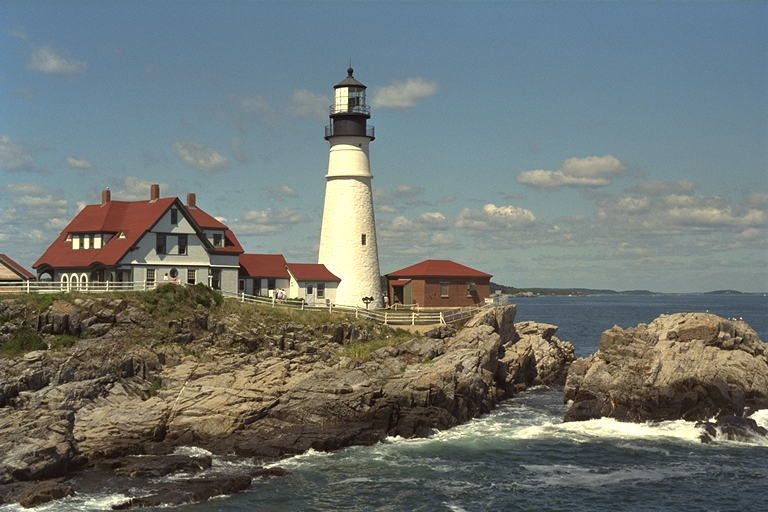}}

\hspace{-4pt}
\raisebox{0.2in}{\rotatebox{90}{$q=0$}}
\stackunder[2pt]{\includegraphics[trim=150 50 50 40,clip,width = 1.09in]{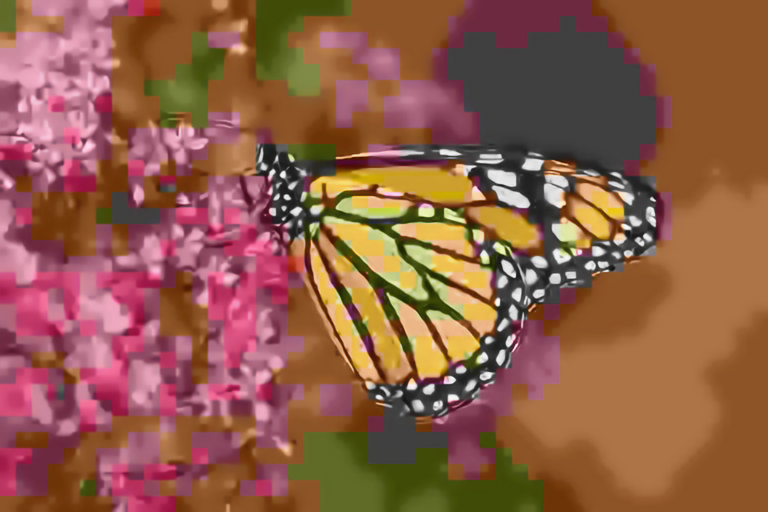}}{DnCNN \cite{dncnn}}
\stackunder[2pt]{\includegraphics[trim=150 50 50 40,clip,width = 1.09in]{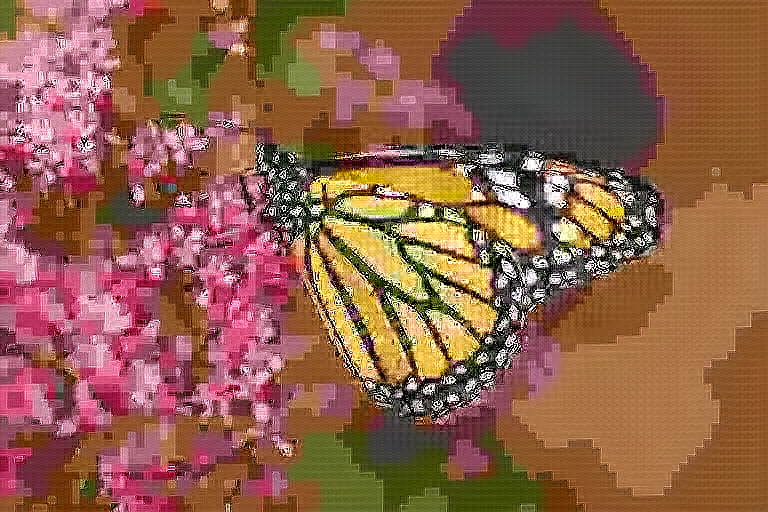}}{QGAC \cite{qgac}}
\stackunder[2pt]{\includegraphics[trim=150 50 50 40,clip,width = 1.09in]{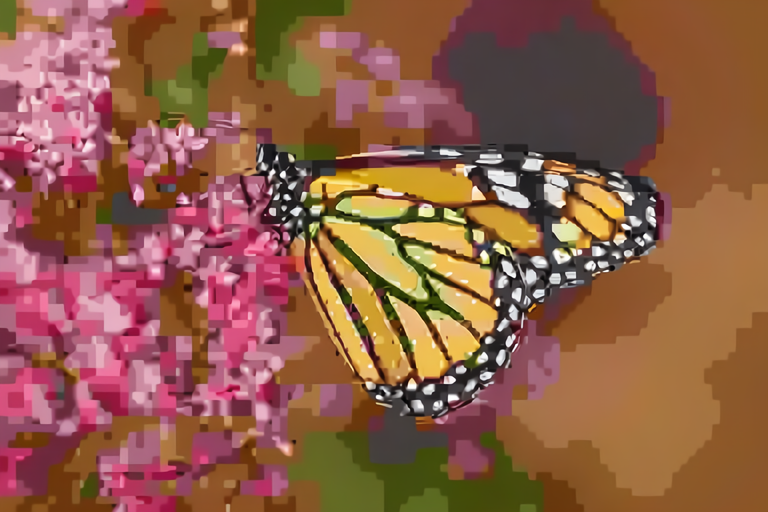}}{FBCNN \cite{fbcnn}}
\stackunder[2pt]{\includegraphics[trim=150 50 50 40,clip,width = 1.09in]{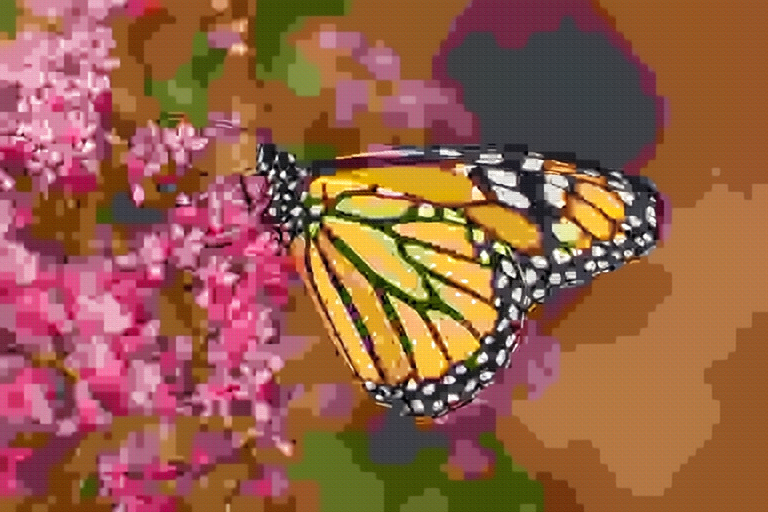}}{JDEC \cite{jdec2024han}}
\stackunder[2pt]{\includegraphics[trim=150 50 50 40,clip,width = 1.09in]{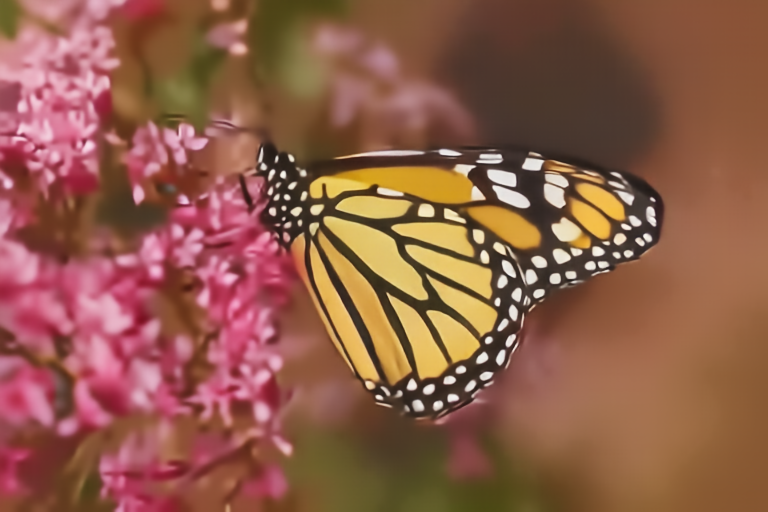}}{\textbf{JPNeO}}
\stackunder[2pt]{\includegraphics[trim=150 50 50 40,clip,width = 1.09in]{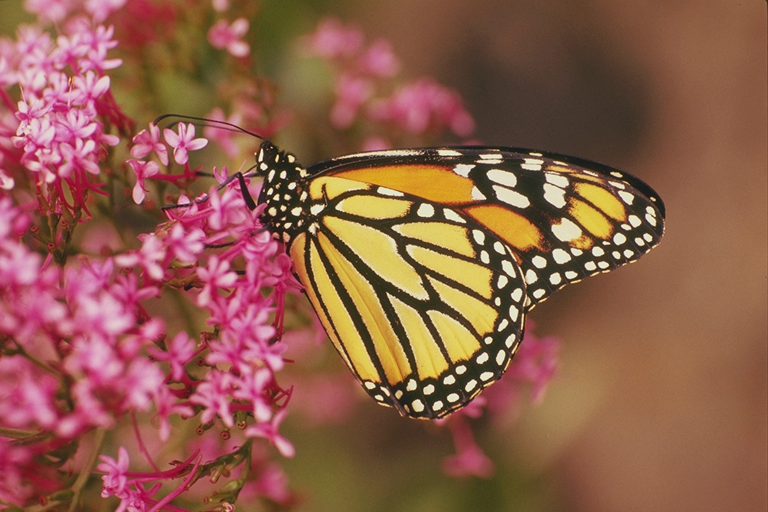}}{GT}
\vspace*{-10pt}
\caption{\textbf{Qualitative Comparison} in color JPEG artifact removal against \textit{JPEG artifact removal networks} ($q=5$ (top), $q=0$ (bottom)).}
\vspace*{-10pt}
\label{fig:qual_comp_main}
\end{figure*}

\begin{figure*}[t]
\footnotesize
\centering
\hspace{-5pt}
\raisebox{0.2in}{\rotatebox{90}{$q=10$}}
{\includegraphics[trim=0 320 550 40,clip,width = 1.09in]{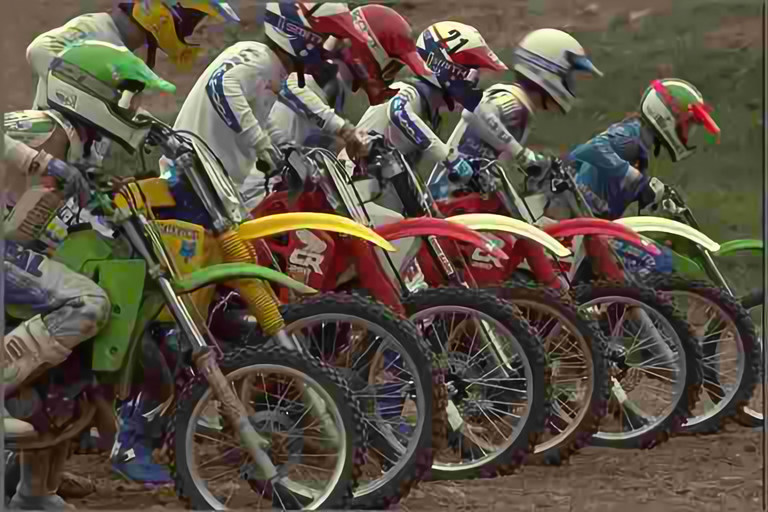}}
{\includegraphics[trim=0 320 550 40,clip,width = 1.09in]{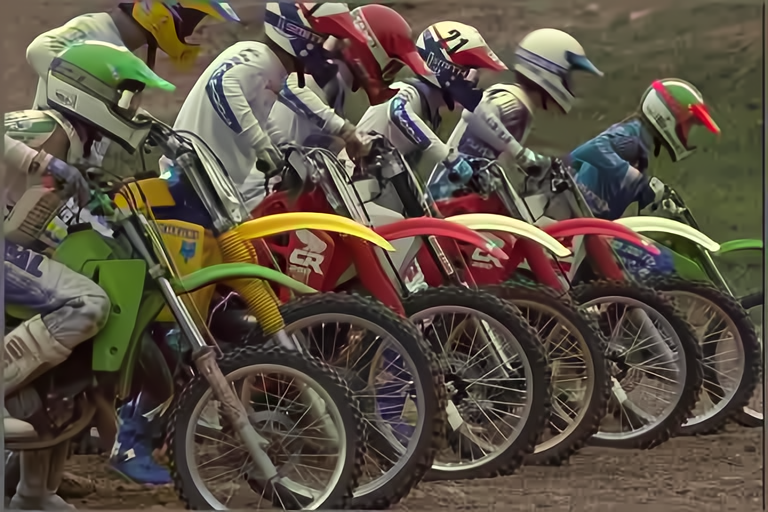}}
{\includegraphics[trim=0 320 550 40,clip,width = 1.09in]{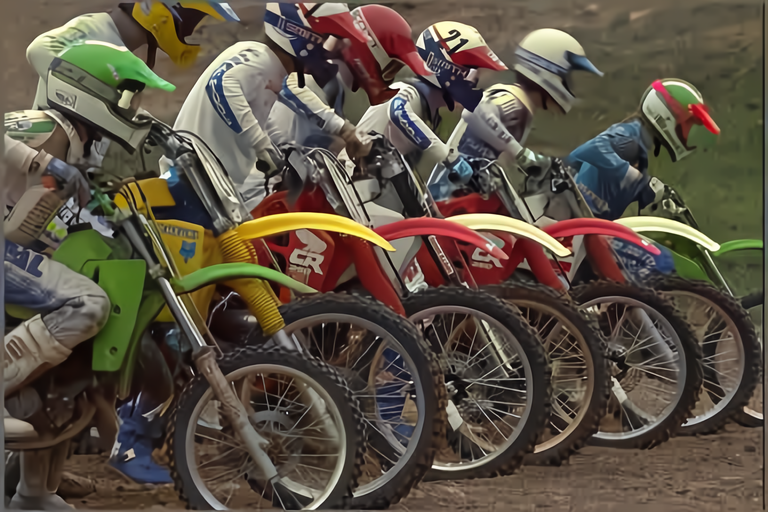}}
{\includegraphics[trim=0 320 550 40,clip,width = 1.09in]{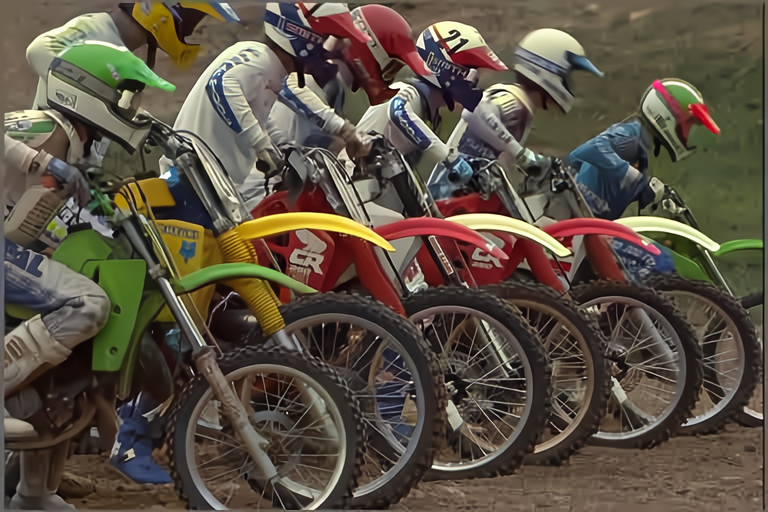}}
{\includegraphics[trim=0 320 550 40,clip,width = 1.09in]{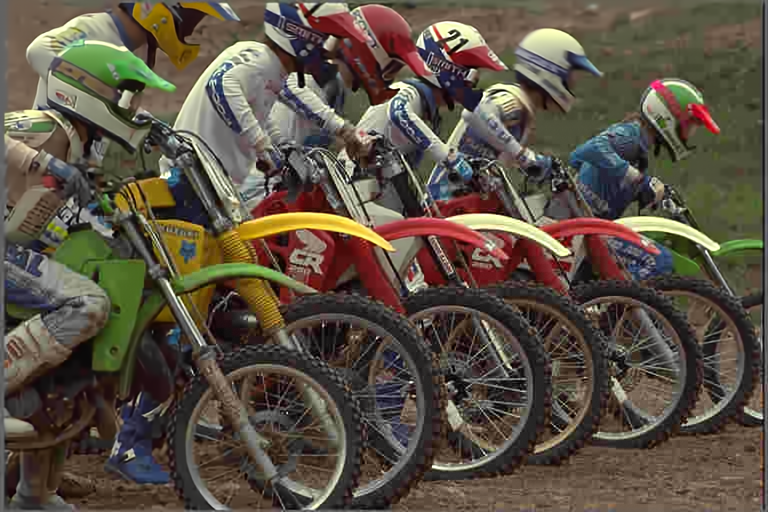}}
{\includegraphics[trim=0 320 550 40,clip,width = 1.09in]{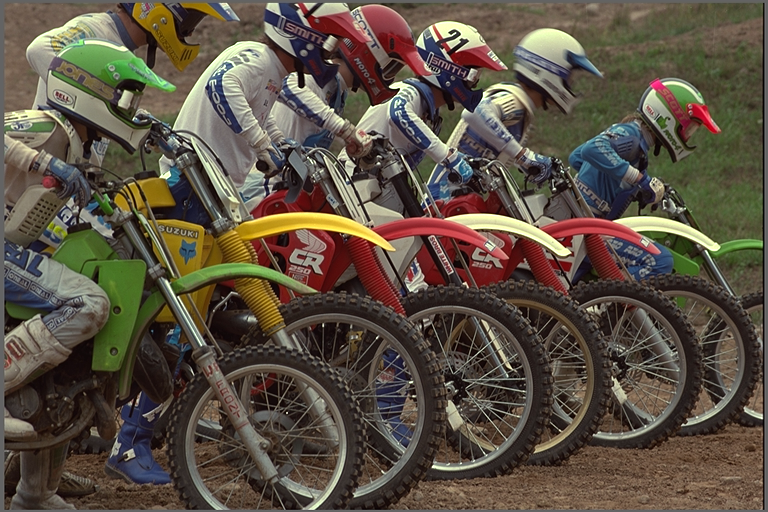}}

\raisebox{0.2in}{\rotatebox{90}{Chroma}}
\stackunder[2pt]{\includegraphics[trim=0 320 550 40,clip,width = 1.09in]{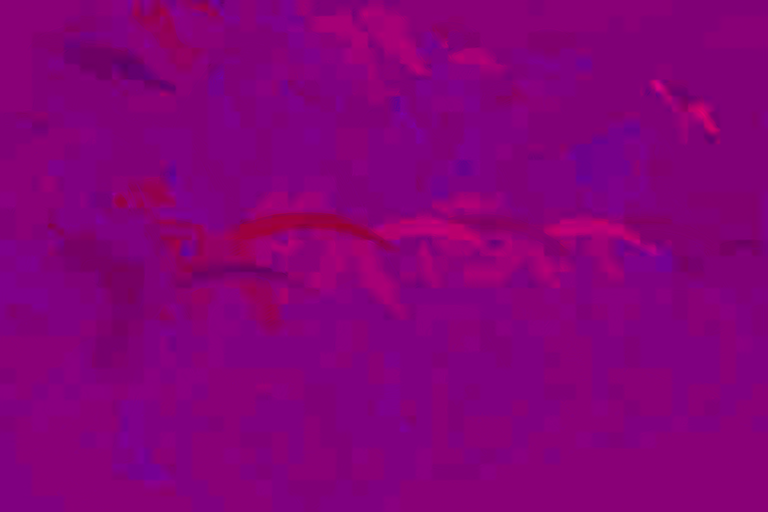}}{DnCNN \cite{dncnn}}
\stackunder[2pt]{\includegraphics[trim=0 320 550 40,clip,width = 1.09in]{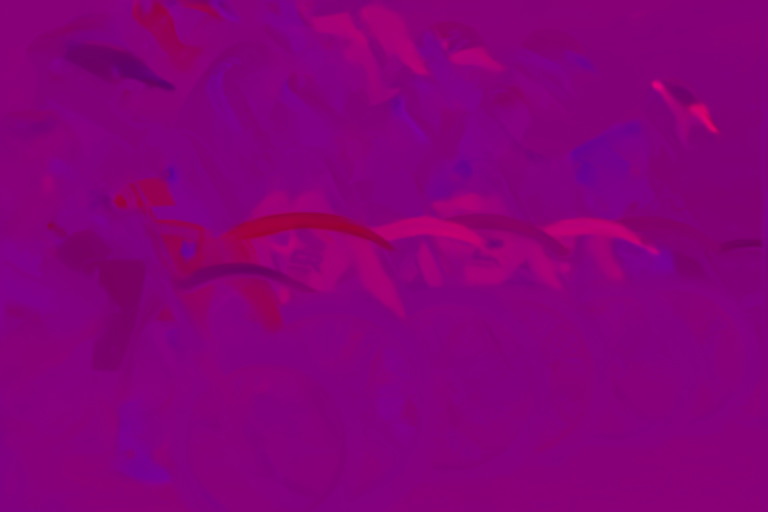}}{QGAC \cite{qgac}}
\stackunder[2pt]{\includegraphics[trim=0 320 550 40,clip,width = 1.09in]{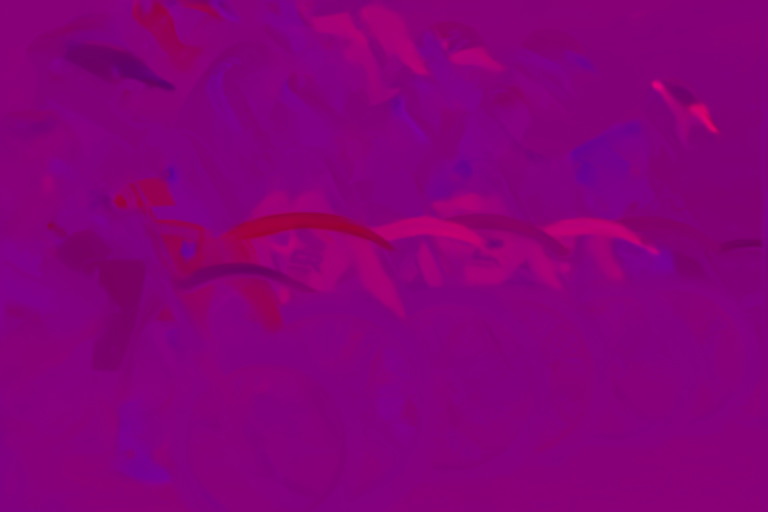}}{FBCNN \cite{fbcnn}}
\stackunder[2pt]{\includegraphics[trim=0 320 550 40,clip,width = 1.09in]{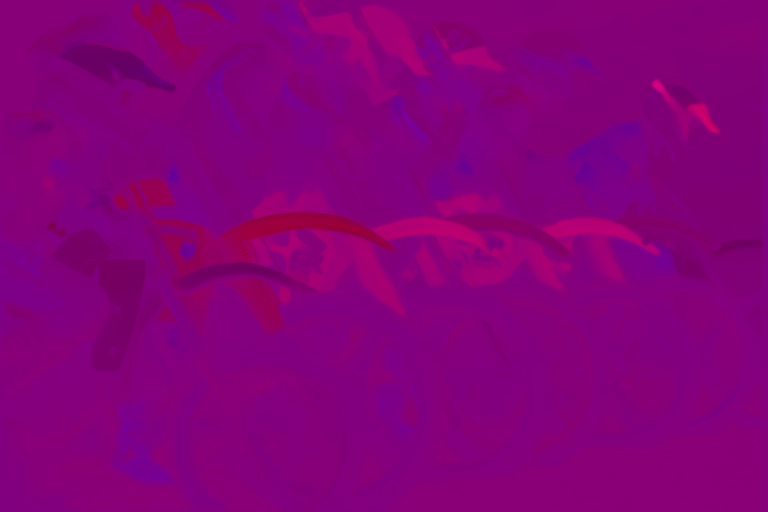}}{JDEC \cite{jdec2024han}}
\stackunder[2pt]{\includegraphics[trim=0 320 550 40,clip,width = 1.09in]{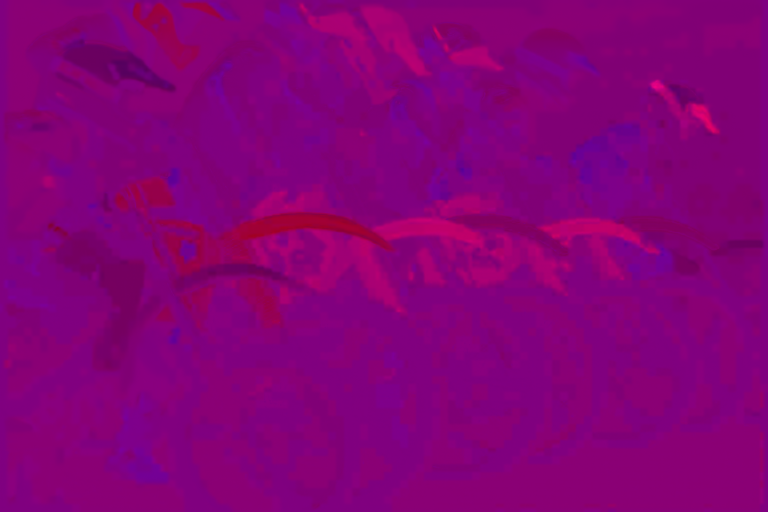}}{\textbf{JPNeO}}
\stackunder[2pt]{\includegraphics[trim=0 320 550 40,clip,width = 1.09in]{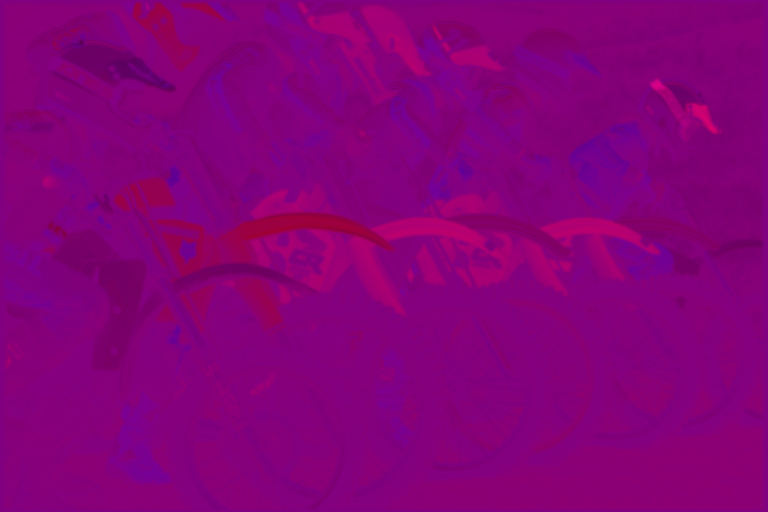}}{GT}
\vspace{-10pt}
\caption{\textbf{Qualitative Comparison} in color JPEG artifact removal with chroma components ($q=10$).}
\vspace*{-10pt}
\label{fig:chroma_cop}
\end{figure*}

\subsection{Training}

\noindent \textbf{Dataset}
For training JPNeO, we use DIV2K and Flickr2K \cite{DIV2kdataset} following the previous work \cite{qgac,fbcnn,jdec2024han} which contains 800 and 2,650 images, respectively. We crop images with size of $112\times112$ which is the least common multiple of JPEG ($=16$ by considering subsampling) and the window size of Swin architecture ($=7$) \cite{conde2022swin2sr,jdec2024han,rgbnomore}. For JENO, we randomly choose a chroma subsampling mode between 4:2:0 and 4:2:2. For JDNO, we combine standard quantization matrix $\mathbf{Q}$ with steps of 10 in the range [10,90] and pre-trained $[\mathbf{Q}_\psi]$ and randomly sample one.


\noindent \textbf{Implementation Detail} 
We use 4 GPUs (NVIDIA RTX 3090 24GB) for training.
Our JPNeO is optimized by Adam \cite{Adamoptimizer} for 1,000 epochs with batch size 64. 
The learning rate is initialized as 1e-4 and decayed by factor 0.5 at $[200,400,600,800]$.



\subsection{Evaluation}\label{sec:abl}

\noindent\textbf{Module-Wise Evaluation} We conduct an evaluation on each part of JPNeO: JENO and JDNO.
We use LIVE1\cite{Live1} and the test set of BSDS500 \cite{B500dataset}.
 In \cref{fig:rd_self}, we fix the encoder or decoder as ($E_\text{JPEG},D_\text{JPEG}$) and measure the performance of our JENO and JDNO, respectively. The performance gain is more pronounced in JDNO at low bpp, whereas JENO shows greater improvement at high bpp. In other words, JENO enhances the upper bound of image quality, while JDNO raises the lower bound, and $\mathbf{Q},\mathbf{Q}_\psi$ determines the path between them.
 To demonstrate this, we highlight the high bpp region for JENO and the low bpp region for JDNO, and present the corresponding SSIM results in \cref{fig:rd_self}. 
In \cref{fig:Qual_self_jpeg} we demonstrate the qualitative analysis for our JPNeO. The performance of JDNO depends on the quality of the encoded image. Using both JENO and JDNO reduces both file size and distortion.
 In \cref{fig:qual_chroma_abl_self}, a qualitative comparison of chroma components at high bpp demonstrates that JENO is robust to chroma subsampling. 
 Regardless of $\mathbf{Q}$, JENO preserves chroma components better than the conventional JPEG encoder, achieving higher performance at lower bpp.

\noindent \textbf{Quantitative Result} To validate our JPNeO, we compare JPNeO with existing compression artifact removal models: DnCNN \cite{dncnn}, QGAC \cite{qgac}, FBCNN \cite{fbcnn}, and JDEC \cite{jdec2024han}. JDEC \cite{jdec2024han} is designed to directly decode spectra into JPEG images, whereas the others act as auxiliary decoders for $D_\text{JPEG}$ covering a wide range of quality factors with a single model. For fair comparison, we evaluate JPNeO using the standard quantization matrix $\mathbf{Q}$ instead of $\mathbf{Q}_\psi$ and 4:2:0 chroma subsampling. Consistent with previous works, we use LIVE-1 \cite{Live1} and the test set from BSDS500 \cite{B500dataset}. As shown in \cref{fig:RDcurve-qual}, JPNeO achieves superior performance compared to all other methods, even at extremely low bpp ($\leq0.5$).

\begin{table}[t]
\centering
\setlength{\tabcolsep}{1.2pt}
\scriptsize
\begin{tabular}{c|c|c c c c c}
\hline
\multicolumn{2}{c|}{{Methods}} & {JPEG}\cite{jpegstandard} & {QGAC\cite{qgac}} & {FBCNN}\cite{fbcnn} & {JDEC}\cite{jdec2024han} & \textbf{JPNeO}\\
\cline{1-2}
{Dataset} & \textbf{$q$} & \#Params. & {259.4M} & {70.1M} & { 38.9M} & \textbf{29.7M}\\
\hline
\multirow{10}{*}{LIVE-1 \cite{Live1}}
& \multirow{2}{*}{0}
& 20.89$|$19.73
& 16.33$|$15.99
& \textcolor{blue}{21.70}$|$\textcolor{blue}{21.19}
& \textcolor{black}{20.76}$|$\textcolor{black}{20.07}
& \textcolor{red}{23.15}$|$\textcolor{red}{22.64} \\
&
& \textcolor{black}{0.540}
& 0.318
& \textcolor{blue}{0.577}
& \textcolor{black}{0.542}
& \textcolor{red}{0.631} \\

& \multirow{2}{*}{10}
& 25.69$|$24.20
& 27.65$|$27.43
& 27.77$|$27.51
& \textcolor{blue}{27.95}$|$\textcolor{red}{27.71}
& \textcolor{red}{28.15}$|$\textcolor{blue}{27.55} \\
&
& 0.759
& 0.819
& 0.816
& \textcolor{blue}{0.821}
& \textcolor{red}{0.829} \\

& \multirow{2}{*}{20}
& 28.06$|$26.49
& 29.88$|$29.56
& 30.11$|$\textcolor{blue}{29.70}
& \textcolor{blue}{30.26}$|$\textcolor{red}{29.87}
& \textcolor{red}{30.49}$|$\textcolor{black}{29.69} \\
&
& 0.841
& 0.882
& 0.881
& \textcolor{blue}{0.885}
& \textcolor{red}{0.890} \\

& \multirow{2}{*}{30}
& 29.37$|$27.84
& 31.17$|$30.77
& 31.43$|$\textcolor{black}{30.92}
& \textcolor{blue}{31.59}$|$\textcolor{red}{31.12}
& \textcolor{red}{31.83}$|$\textcolor{blue}{30.94} \\
&
& 0.875
& 0.908
& \textcolor{black}{0.908}
& \textcolor{blue}{0.911}
& \textcolor{red}{0.915} \\

& \multirow{2}{*}{40}
& 30.28$|$28.84
& 32.08$|$31.64
& 32.34$|$\textcolor{black}{31.80}
& \textcolor{blue}{32.50}$|$\textcolor{red}{31.98}
& \textcolor{red}{32.83}$|$\textcolor{blue}{31.91} \\
&
& 0.894
& 0.922
& \textcolor{black}{0.923}
& \textcolor{blue}{0.925}
& \textcolor{red}{0.929} \\
\hline

\multirow{10}{*}{BSDS500 \cite{B500dataset}}
& \multirow{2}{*}{0}
& 20.82$|$19.57
& 16.61$|$16.17
& \textcolor{blue}{21.63}$|$ \textcolor{blue}{21.08}
& \textcolor{black}{20.63}$|$\textcolor{black}{19.89}
& \textcolor{red}{23.10}$|$\textcolor{red}{22.58} \\
&
& 0.517
& 0.314
& \textcolor{blue}{0.558}
& \textcolor{black}{0.514}
& \textcolor{red}{0.618} \\

& \multirow{2}{*}{10}
& 25.84$|$24.13
& 27.75$|$27.48
& 27.85$|$\textcolor{blue}{27.53}
& \textcolor{blue}{28.00}$|$\textcolor{red}{27.67}
& \textcolor{red}{28.06}$|$\textcolor{black}{27.29} \\
&
& 0.759
& 0.819
& 0.814
& \textcolor{blue}{0.819}
& \textcolor{red}{0.824} \\

& \multirow{2}{*}{20}
& 28.21$|$26.37
& 30.04$|$29.55
& 30.14$|$29.58
& \textcolor{blue}{30.31}$|$\textcolor{red}{29.71}
& \textcolor{red}{30.36}$|$\textcolor{black}{29.31} \\
&
& 0.844
& 0.884
& 0.881
& \textcolor{blue}{0.885}
& \textcolor{red}{0.887} \\

& \multirow{2}{*}{30}
& 29.57$|$27.72
& 31.36$|$30.73
& 31.45$|$\textcolor{blue}{30.74}
& \textcolor{blue}{31.65}$|$\textcolor{red}{30.88}
& \textcolor{red}{31.70}$|$\textcolor{black}{30.49} \\
&
& 0.880
& 0.911
& 0.909
& \textcolor{blue}{0.912}
& \textcolor{red}{0.914} \\

& \multirow{2}{*}{40}
& 30.52$|$28.69
& 32.29$|$31.53
& 32.36$|$31.54
& \textcolor{blue}{32.53}$|$\textcolor{red}{31.68}
& \textcolor{red}{32.70}$|$\textcolor{black}{31.40} \\
&
& 0.900
& 0.926
& 0.924
& \textcolor{blue}{0.927}
& \textcolor{red}{0.929} \\
\hline

\end{tabular}
\vspace{-10pt}
\caption{Quantitative comparisons (PSNR (dB) $|$ PSNR-B (dB) (top), SSIM (bottom)) with \textit{the JPEG artifact removal} networks. Note that the JPEG artifact removal networks are auxiliary decoder of the JPEG. We highlight the best and the second-best performance with \textcolor{red}{red} and \textcolor{blue}{blue} colors, respectively.}
\vspace{-20pt}
\label{tab:quantitative_main}
\end{table}

For a numerical comparison, we report PSNR, structural similarity index (SSIM), and PSNR-B against baselines with the number of parameters in \cref{tab:quantitative_main}.
We compare the existing artifact removal networks with quality factors \{0, 10, 20, 30, 40\} for both datasets.
As demonstrated in \cref{tab:quantitative_main}, JPNeO demonstrates superior performance in terms of PSNR and SSIM when compared to existing networks. Notably, the performance of the proposed model is particularly pronounced in quality factor 0.
We report PSNR and SSIM of each luminance and chroma component in \cref{tab:chroma-comparison} for a detailed comparison. The results demonstrate that the enhanced performance of JPNeO is a consequence of its ability to restore chroma components more effectively than other methods.

\begin{table}[t]
\centering
\setlength{\tabcolsep}{1.2pt}
\scriptsize
\begin{tabular}{c|c|c c c c c}
\hline
\multicolumn{2}{c|}{{Methods}} & \multirow{2}{*}{DnCNN \cite{dncnn}} & \multirow{2}{*}{QGAC \cite{qgac}} & \multirow{2}{*}{FBCNN \cite{fbcnn}} & \multirow{2}{*}{JDEC \cite{jdec2024han}} & \multirow{2}{*}{\textbf{JPNeO}}\\
\cline{1-2}
{Dataset} & \textbf{$q$} &  &  &  &  & \\
\hline
\multirow{10}{*}{LIVE-1 \cite{Live1}} 
& \multirow{2}{*}{0}
& \textcolor{blue}{23.80}$|$\textcolor{black}{29.27}
& \textcolor{black}{16.76}$|$\textcolor{black}{29.61}
& \textcolor{black}{23.23}$|$\textcolor{blue}{29.86}
& \textcolor{black}{22.89}$|$\textcolor{black}{28.75}
& \textcolor{red}{24.20}$|$\textcolor{red}{32.30} \\
& 
& \textcolor{blue}{0.304}$|$\textcolor{black}{0.044}
& \textcolor{black}{0.135}$|$\textcolor{black}{0.039}
& \textcolor{black}{0.286}$|$\textcolor{blue}{0.068}
& \textcolor{black}{0.270}$|$\textcolor{black}{0.054}
& \textcolor{red}{0.342}$|$\textcolor{red}{0.125} \\

& \multirow{2}{*}{10}
& \textcolor{black}{28.39}$|$\textcolor{black}{34.47}
& \textcolor{black}{28.70}$|$\textcolor{black}{37.02}
& \textcolor{black}{28.76}$|$\textcolor{black}{37.35}
& \textcolor{red}{28.83}$|$\textcolor{blue}{37.95}
& \textcolor{blue}{28.77}$|$\textcolor{red}{38.56} \\
& 
& \textcolor{black}{0.582}$|$\textcolor{black}{0.101}
& \textcolor{black}{0.602}$|$\textcolor{black}{0.206}
& \textcolor{black}{0.600}$|$\textcolor{black}{0.255}
& \textcolor{blue}{0.604}$|$\textcolor{blue}{0.274}
& \textcolor{red}{0.616}$|$\textcolor{red}{0.291} \\

& \multirow{2}{*}{20}
& \textcolor{black}{30.77}$|$\textcolor{black}{37.15}
& \textcolor{black}{31.01}$|$\textcolor{black}{39.01}
& \textcolor{blue}{31.13}$|$\textcolor{black}{39.64}
& \textcolor{red}{31.16}$|$\textcolor{blue}{40.25}
& \textcolor{black}{30.88}$|$\textcolor{red}{41.52} \\
&
& \textcolor{black}{0.697}$|$\textcolor{black}{0.175}
& \textcolor{black}{0.708}$|$\textcolor{black}{0.281}
& \textcolor{black}{0.707}$|$\textcolor{black}{0.330}
& \textcolor{blue}{0.711}$|$\textcolor{blue}{0.355}
& \textcolor{red}{0.713}$|$\textcolor{red}{0.391} \\

& \multirow{2}{*}{30}
& \textcolor{black}{32.14}$|$\textcolor{black}{38.27}
& \textcolor{black}{32.40}$|$\textcolor{black}{40.20}
& \textcolor{blue}{32.53}$|$\textcolor{black}{40.63}
& \textcolor{red}{32.55}$|$\textcolor{blue}{41.31}
& \textcolor{black}{32.18}$|$\textcolor{red}{42.72} \\
&
& \textcolor{black}{0.748}$|$\textcolor{black}{0.221}
& \textcolor{black}{0.756}$|$\textcolor{black}{0.324}
& \textcolor{black}{0.756}$|$\textcolor{black}{0.372}
& \textcolor{blue}{0.758}$|$\textcolor{blue}{0.402}
& \textcolor{red}{0.759}$|$\textcolor{red}{0.452} \\

& \multirow{2}{*}{40}
& \textcolor{black}{33.14}$|$\textcolor{black}{38.98}
& \textcolor{blue}{33.37}$|$\textcolor{black}{40.79}
& \textcolor{red}{33.53}$|$\textcolor{black}{41.23}
& \textcolor{red}{33.53}$|$\textcolor{blue}{41.92}
& \textcolor{black}{33.18}$|$\textcolor{red}{43.47} \\
&
& \textcolor{black}{0.776}$|$\textcolor{black}{0.256}
& \textcolor{black}{0.782}$|$\textcolor{black}{0.351}
& \textcolor{black}{0.783}$|$\textcolor{black}{0.399}
& \textcolor{blue}{0.785}$|$\textcolor{blue}{0.431}
& \textcolor{red}{0.785}$|$\textcolor{red}{0.494} \\

\hline
\multirow{10}{*}{BSDS500 \cite{B500dataset}} 
& \multirow{2}{*}{0}
& \textcolor{blue}{23.22}$|$\textcolor{black}{28.42}
& \textcolor{black}{16.72}$|$\textcolor{black}{28.78}
& \textcolor{black}{22.70}$|$\textcolor{blue}{29.18}
& \textcolor{black}{21.96}$|$\textcolor{black}{27.84}
& \textcolor{red}{23.72}$|$\textcolor{red}{31.79} \\
& 
& \textcolor{blue}{0.306}$|$\textcolor{black}{0.058}
& \textcolor{black}{0.141}$|$\textcolor{black}{0.045}
& \textcolor{black}{0.294}$|$\textcolor{blue}{0.091}
& \textcolor{black}{0.271}$|$\textcolor{black}{0.035}
& \textcolor{red}{0.360}$|$\textcolor{red}{0.164} \\

& \multirow{2}{*}{10}
& \textcolor{black}{27.50}$|$\textcolor{black}{34.91}
& \textcolor{black}{27.72}$|$\textcolor{black}{37.70}
& \textcolor{black}{27.75}$|$\textcolor{black}{37.81}
& \textcolor{blue}{27.78}$|$\textcolor{blue}{38.42}
& \textcolor{red}{27.89}$|$\textcolor{red}{39.01} \\
& 
& \textcolor{black}{0.582}$|$\textcolor{black}{0.101}
& \textcolor{blue}{0.636}$|$\textcolor{black}{0.264}
& \textcolor{black}{0.630}$|$\textcolor{black}{0.302}
& \textcolor{black}{0.634}$|$\textcolor{blue}{0.321}
& \textcolor{red}{0.649}$|$\textcolor{red}{0.138} \\

& \multirow{2}{*}{20}
& \textcolor{black}{29.72}$|$\textcolor{black}{37.99}
& \textcolor{black}{29.92}$|$\textcolor{black}{40.36}
& \textcolor{blue}{29.98}$|$\textcolor{black}{40.41}
& \textcolor{red}{30.00}$|$\textcolor{blue}{41.02}
& \textcolor{black}{29.95}$|$\textcolor{red}{42.64} \\
&
& \textcolor{black}{0.740}$|$\textcolor{black}{0.218}
& \textcolor{black}{0.753}$|$\textcolor{black}{0.346}
& \textcolor{black}{0.751}$|$\textcolor{black}{0.376}
& \textcolor{blue}{0.755}$|$\textcolor{blue}{0.407}
& \textcolor{red}{0.758}$|$\textcolor{red}{0.448} \\

& \multirow{2}{*}{30}
& \textcolor{black}{31.12}$|$\textcolor{black}{39.30}
& \textcolor{black}{31.31}$|$\textcolor{black}{41.36}
& \textcolor{blue}{31.39}$|$\textcolor{black}{41.39}
& \textcolor{red}{31.41}$|$\textcolor{blue}{42.15}
& \textcolor{black}{31.27}$|$\textcolor{red}{44.11} \\
&
& \textcolor{black}{0.797}$|$\textcolor{black}{0.270}
& \textcolor{black}{0.806}$|$\textcolor{black}{0.389}
& \textcolor{black}{0.805}$|$\textcolor{black}{0.409}
& \textcolor{blue}{0.808}$|$\textcolor{blue}{0.444}
& \textcolor{red}{0.809}$|$\textcolor{red}{0.508} \\

& \multirow{2}{*}{40}
& \textcolor{black}{32.14}$|$\textcolor{black}{40.23}
& \textcolor{black}{32.32}$|$\textcolor{black}{42.19}
& \textcolor{blue}{32.41}$|$\textcolor{black}{42.06}
& \textcolor{red}{32.43}$|$\textcolor{blue}{42.81}
& \textcolor{black}{32.32}$|$\textcolor{red}{45.20} \\
&
& \textcolor{black}{0.829}$|$\textcolor{black}{0.315}
& \textcolor{black}{0.836}$|$\textcolor{black}{0.422}
& \textcolor{black}{0.836}$|$\textcolor{black}{0.436}
& \textcolor{blue}{0.837}$|$\textcolor{blue}{0.472}
& \textcolor{red}{0.839}$|$\textcolor{red}{0.567} \\

\hline
\end{tabular}
\vspace{-10pt}
\caption{{Quantitative comparisons of luminance and chroma components. ($\mathbf{X}_Y$-PSNR$|\mathbf{X}_C$-PSNR (top),$\mathbf{X}_Y$-SSIM$|\mathbf{X}_C$-SSIM (bottom))}}\label{tab:chroma-comparison}
\vspace{-20pt}
\end{table}

\noindent \textbf{Qualitative Result} We show qualitative results against JPEG artifact removal networks with low quality factors \{0, 5\} in \cref{fig:qual_comp_main}.
While other networks do not fully remove JPEG artifacts with low-quality factors, our JPNeO significantly removes the artifacts such as the monarch of the second row of \cref{fig:qual_comp_main} maintaining the color tones.
In \cref{fig:chroma_cop}, the comparison shows qualitative results in the RGB and chroma domains with quality factor 10. While the results in the RGB domain show a slight perceptual difference, the results in the chroma components show that JPNeO restores more fine details than the other networks.

\section{Discussion}
\label{sec:discussuin}
\begin{table}[t]
    \centering
    \scriptsize
    \setlength{\tabcolsep}{1.2pt}
    \begin{tabular}{c|c|c|c|c|c|c|c|c}
       {PSNR (dB)}   &\#Params.& {Mem.} & {Time} &MACs & \multicolumn{2}{c|}{LIVE1 \cite{Live1}}&\multicolumn{2}{c}{BSDS500 \cite{B500dataset}} \\
        \cline{1-1} \cline{6-9}
         Method &(M)&(GB)&(ms)& (K/Pixel) &$q=0$&$q=40$&$q=0$&$q=40$\\ 
        \hline\hline
         FBCNN \cite{fbcnn}  & 70.1 &  {0.61} & \textcolor{red}{71.95} &\textcolor{blue}{1131}&21.70&32.34&21.63&32.36 \\
         JDEC \cite{jdec2024han} & 38.9 &  {1.76} &  {224.79}&  {1605} &20.76  &32.50 &20.63&\textcolor{black}{32.53}\\
             \hline
     JPNeO$^-$ &  \textcolor{red}{8.0} & \textcolor{red}{0.09}          &  \textcolor{blue}{222.95}&  \textcolor{red}{1045} &\textcolor{blue}{22.98} &\textcolor{blue}{32.72} &\textcolor{blue}{22.73}&\textcolor{blue}{32.63}\\
      JPNeO& \textcolor{blue}{29.7} &  \textcolor{blue}{0.26} &  562.42&  {3489} &\textcolor{red}{23.15} & \textcolor{red}{32.83}&\textcolor{red}{23.10}&\textcolor{red}{32.70} \\
    \end{tabular}
    \vspace*{-6pt}
    \caption{Quantitative comparison of computational costs \& decoding performance.}
    \label{tab:comp_table}
    \vspace{-7pt}
\end{table}

\begin{table}[t]
\vspace{-10pt}
    \centering
    \scriptsize
    \setlength{\tabcolsep}{1.2pt}
    \begin{tabular}{c|c|c|c||c|c|c|c||c
            }
        &\multicolumn{3}{c||}{JENO ($E_\varphi$)}&\multicolumn{4}{c||}{JDNO ($D_\theta$)}&\multirow{2}{*}{JPNeO}\\
        \cline{2-8}
          &$f_\xi$&$\mathcal{G}_{\phi}$& Total & $f_\xi$&$T_\rho$&$\mathcal{G}_{\phi}$& Total& \\ 
        \hline\hline
\#Params. (M) &1.220 &1.663 &2.883 &22.72 &0.902 &3.16 &26.78 &29.663 \\
MACs (K/Pixel) & 609 &541 &1150& 732&10&1664 &2339 &3489\\
\hline
\hline
Time (ms)\tablefootnote{Per-module measurement are excluded due to GPU kernel overlap.} & \multicolumn{3}{c||}{177.85} &\multicolumn{4}{c||}{398.8}&562.42\\
Mem. (MB)& \multicolumn{3}{c||}{26.85} &\multicolumn{4}{c||}{189.81}&246.62\\
    \end{tabular}
    \vspace*{-6pt}
    \caption{Module-wise resource consumption (parameters, MACs, and time\& memory consumption) within the encoder and decoder.}
    \label{tab:comp_table_modulo}
    \vspace{-10pt}
\end{table}

\begin{figure}[t]
\footnotesize
\centering
\includegraphics[trim= 0 0 0 0,clip,width = 3.0in]{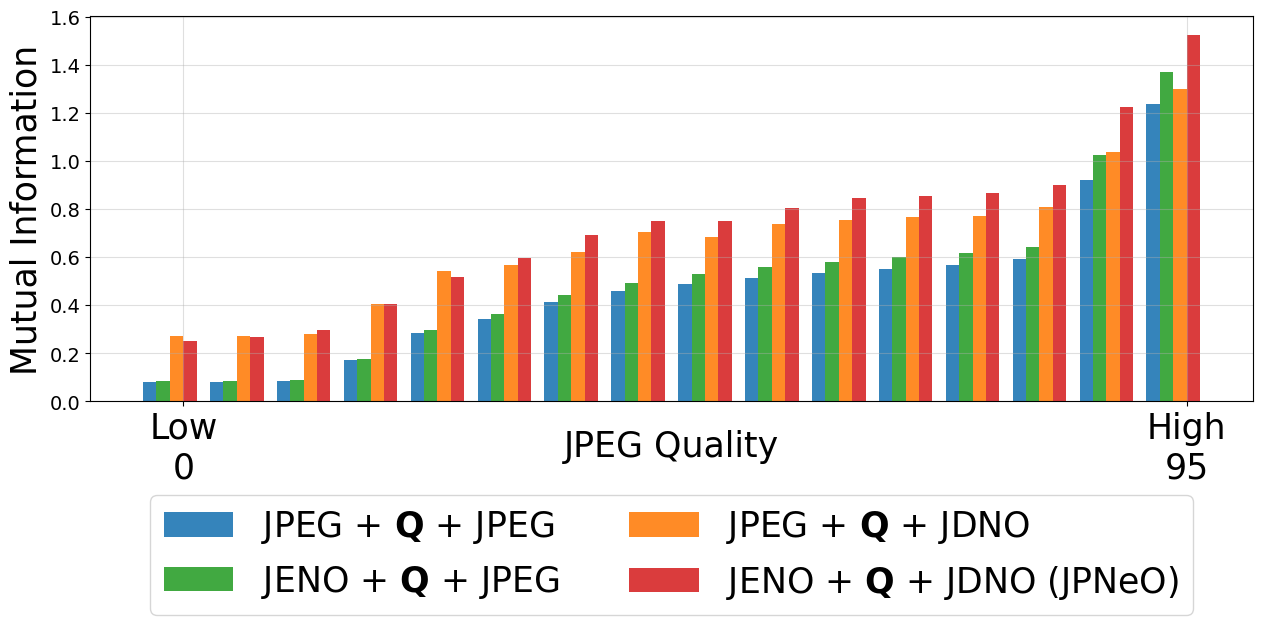}
\vspace{-10pt}
\caption{\textbf{Mutual information} with the ground-truth image ($I(\mathbf{X};\mathbf{\hat X})$) across quality levels in Kodak \cite{Kodakdataset} (4:2:0 subsampling). The label indicates `Encoder ($E_\varphi/E_\text{JPEG}$)$+$`$\mathbf{Q/Q}_\psi$'$+$Decoder ($D_\theta/D_\text{JPEG}$)' }
\label{fig:mutual_info}
\vspace{-18pt}
\end{figure}



\noindent\textbf{Computational Costs} In \cref{tab:comp_table,tab:comp_table_modulo}, we report computational cost and performance, including network parameters, time and memory consumption, and Multiply–Accumulates (MACs) per pixel. We further includes a lightweight variant (JPNeO$^{-}$) evaluated under similar MAC/pixel constraints for a fair comparison.
The size of the input for the comparison is $560\times 560$. 
Notably, our JPNeO and JPNeO$^-$ surpass the performance of FBCNN\cite{fbcnn} and JDEC\cite{jdec2024han} while requiring only a minimal amount of resources which is a significant outcome considering the adoption of JPEG across ISP pipelines.
In \cref{tab:comp_table_modulo}, we decompose each model to compare parameter counts and MACs at the module level.



\begin{figure}[!t]
\footnotesize
\centering
\includegraphics[trim= 0 0 0 0,clip,width = 3.0in]{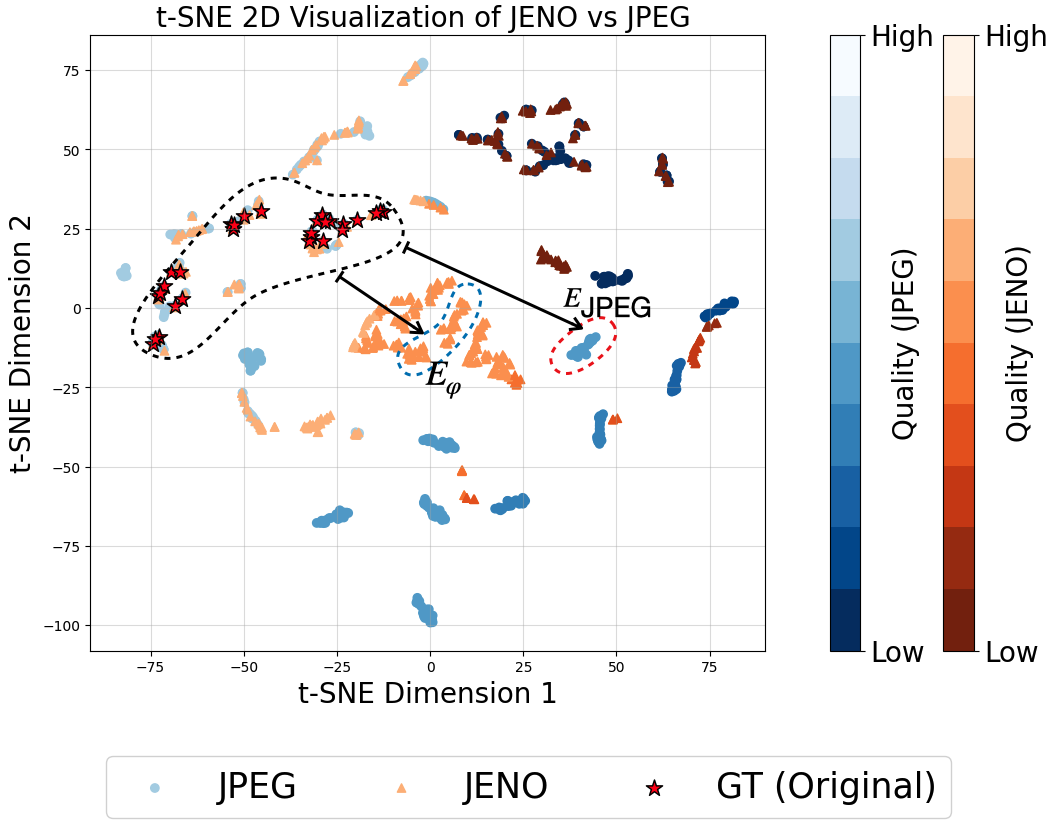}
\vspace{-10pt}
\caption{\textbf{t-SNE \cite{van2008visualizingtsne} clustering analysis} by quality factor on the LIVE-1 \cite{Live1}. Encoded images with our JENO ($E_\varphi$) (\textcolor{red}{reds}) and JPEG encoder ($E_\text{JPEG}$) (\textcolor{blue}{blues}).}
\label{fig:tsne-cluster}
\vspace{-19pt}
\end{figure}

\noindent\textbf{Mutual Information} In \cref{sec:problemformulation} and \cref{eq:dec_jpeg,eq:enc_jpeg}, the mutual information between the prior embedded in the MAP-trained parameters $\theta$ and $\mathbf{\tilde X}$ provides advantages for image restoration. Similarly, the mutual information between $I(\mathbf{X'};\varphi)$ enhances encoding performance. To validate this, we calculate the mutual information using the symbol statistics from the Kodak dataset \cite{Kodakdataset} while switching the encoder and decoder between ($E_\varphi/E_\text{JPEG} + D_\theta/D_\text{JPEG} $). In \cref{fig:mutual_info}, key observation is that as the quality increases, JENO’s mutual information rises, eventually exceeding the mutual information gain achieved by JDNO. However, at low quality, JDNO achieves a larger increase in mutual information. In conclusion, using only JENO at high bpp and only JDNO at low bpp is resource-efficient, while using both at intermediate quality levels is preferable.

\noindent\textbf{t-SNE Clustering} We compare the latent distances when using JENO for compression with those obtained from standard JPEG to support \cref{fig:jpeg_mapping} and our hypothesis in \cref{sec:intro}. 
\cref{fig:tsne-cluster} presents the clustering results after extracting latent vectors using a quality predictor of FBCNN \cite{fbcnn}.
The distance between GT region and the range of $E_\varphi$ is less than the distance between GT region and the range of $E_\text{JPEG}$. \cref{fig:tsne-cluster} show that JPEG compression produces significantly larger distances between latent vectors at the same quality level.
\section{Conclusion}
We propose JPNeO, a learning-based codec that remains backward-compatibility with the existing JPEG format while enhancing compression efficiency and reconstruction quality.
By applying neural operators in both encoding and decoding and training a new quantization matrix, we significantly reduce quantization loss.
Experimental results show that JPNeO successfully restores color components and achieves better PSNR and SSIM compared to existing approaches.
The memory-friendly design and flexible architecture allow seamless integration with legacy JPEG systems. 
Moreover, JPNeO leverages learned image priors to maintain high-fidelity details at both low and high bit rates. 
Thus, it shows strong potential for next-generation JPEG pipelines.
\vspace{8pt}
\noindent
\section*{Acknowledgments}
\small{
This work was partly supported by the National Research Foundation of Korea (NRF) grant funded by the Korea government (MSIT) (RS-2024-00335741, RS-2024-00413303) and Smart Health-Care for Police Officers Program through the Korea Institutes of Police Technology(KIPoT) funded by the Korean National Police Agency (RS-2022-PT000186).}

\noindent

{
    \small
    \bibliographystyle{ieeenat_fullname}
    \bibliography{main}
}

\clearpage
\setcounter{equation}{1}
\setcounter{page}{1}
\renewcommand{\thetable}{S\arabic{table}}  %
\setcounter{table}{0}     
\setcounter{section}{0}
\renewcommand{\thefigure}{S\arabic{figure}}  %
\setcounter{figure}{0}

\twocolumn[{
\renewcommand\twocolumn[1][]{#1}%
\maketitlesupplementary
\begin{center}
    \centering
    \captionsetup{type=figure} 
    \includegraphics[width=1\textwidth]{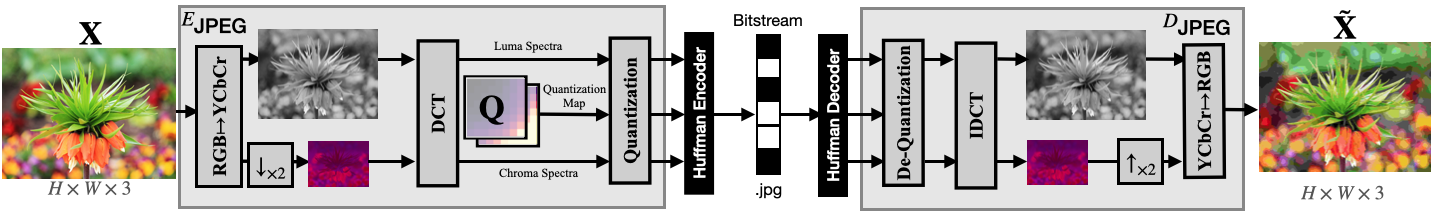}
    \caption{Schematic flow of the standard JPEG compression.}
    \label{fig:jpeg_conventional}
\end{center}%
\begin{center}
    \centering
    \captionsetup{type=figure} 
    \includegraphics[width=1\linewidth]{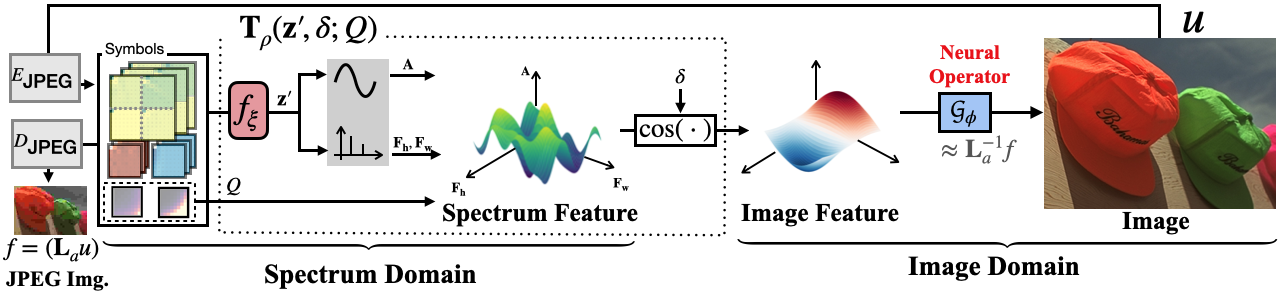}
    \caption{Visual summary of the proposed JPEG decoding. The loss incurred during the JPEG pipeline is modeled as \cref{eq:no_init} of the main paper.}
    \label{fig:dec_process}
\end{center}%
}]

\begin{table*}[ht!]
\scriptsize
    \centering
    \setlength{\tabcolsep}{1.2pt}
    \begin{tabular}{c|l|l|l}
    \toprule
    \textbf{Symbol} & \textbf{Definition} & \textbf{Description} & \textbf{Meaning / Note} \\
    \midrule
    $\mathbf{X}$ & $\in\mathbb{R}^{H\times W\times 3}$ & Original RGB image & Ground-truth \\
    $\mathbf{Y}$ & $=\mathcal{D}_8(\mathbf{X})$ & Real-valued discrete cosine transform's coefficients & \\
    $\mathbf{Q}$ & $\in [1,255]^{8\times 8}$ & Standard JPEG quantization matrix & Integer \\
        $\mathbf{Q}_{\psi}$ & $\in [1,255]^{8\times 8}$ & Learned quant.\ matrix & Learned as float number and stored as integer. \\
    $H,W,K ,M$ & $\in\mathbb{N}$ & Height $(H)$, Width $(W)$, and Channel $(K,M)$ of a Tensor &  \\
        $B$ & $\in\mathbb{N}$ & Block size of the JDNO & $4$ as implementation \\
            $M$ &$\in\mathbb{N}$ & Cosine-feature channels of $T_\rho$ & $128$ as implementation \\
    $r_{1,2}$ & $\in \{0.5,1\}$ & Sub-sampling ratio & $r_{1,2}$ for the height and the width, respectively  \\
    $\delta$ & $\in\mathbb{R}^{H\times W\times2}$ & Coord. grid & Input to JENO and used in $\mathcal{S}$ \\
    $\delta_{Y,C}$ & $\in\mathbb{R}^{r_1H\times r_2W\times2}$ & Coordinates of $\mathbf{z}$ & Used in $\mathcal{S}$ \\
    $\Delta\delta$ & $\in[-1,1]^2 $ & Local coordinates & Used in $\mathcal{S}$ \\
    $s_i$ & $\in \mathbb{R}$ & Local area weight &  Used in $\mathcal{S}$ \\
    $\mathbf{c}$ & $:=(2/r_1,\,2/r_2)$ & Anti-aliasing factor &  Used in $\mathcal{S}$ \\
    $\mathbf{z}$ & $\in \mathbb{R}^{H\times W\times K}$ & Feature map of the JPNeO & \\
    $\lambda$ & $\in ( 0,\infty)$ & Loss trade-off & Used in training $\mathbf{Q_\psi}$ \\
    $\varphi,\psi,\theta,\xi,\rho,\phi$ & – & Trainable parameters of the JPNeO &  \\
    \midrule
    $\mathcal{D}_B$ & $\mathbf{X}\rightarrow \mathbf{Y}$ & Discrete Cosine Transform (DCT) with block-size $B$ & $\mathcal{D}^{-1}$ as an inverse DCT (IDCT) \\
    $E_{\text{JPEG}}$ & $\mathbf{X}\rightarrow \mathbf{Y}'$ & JPEG standard encoder &  \\
    $D_{\text{JPEG}}$ & $ \mathbf{Y}'\rightarrow \tilde{\mathbf{X}}$ & JPEG standard decoder &  \\
    $E_{\varphi}$ & $(\mathbf{X},(\delta_Y,\delta_C))\rightarrow (\mathbf{X}_Y,\mathbf{X}_C)$ & JPEG Encoding Neural Operator (JENO) & $ \mathcal{G}_\phi \circ\mathcal{S}\circ f_\xi $  \\
    $D_{\theta}$ & $(\tilde{\mathbf{Y}}_Y,\tilde{\mathbf{Y}}_C;\mathbf{Q})\rightarrow \hat{\mathbf{X}}$ & JPEG Decoding Neural Operator (JDNO) & $ \mathcal{G}_\phi \circ T_\rho\circ\mathcal{S}\circ f_\xi $ \\
    $f_{\xi}$ & $\mathbf{X}\rightarrow \mathbf{z}$ or $\mathbf{(\tilde{\mathbf{Y}}_Y,\tilde{\mathbf{Y}}_C)}\rightarrow \mathbf{z}$ & Feature extractor of JPNeO & JENO and JDNO, respectively  \\
    $\mathcal{G}_{\varphi}$ & $\mathbf{z} \rightarrow \mathbf{\hat{X}}$ & Neural Operator & Utilizing Galerkin attention \\
    $T_{\rho}$ &$(\mathbf{z}', \delta, \mathbf{Q})\rightarrow \mathbf{z}$ & Cosine Neural Operator (CNO) & $\mathbf{A\otimes(\cos(\pi\mathbf{F}_{h}\otimes\delta_h)\odot \cos(\pi \mathbf{F}_{w}\otimes\delta_w))}$\\
    $h_a,h_f$ &$\mathbb{R}^K \rightarrow \mathbb{R}^M $ & Coefficient and Frequency for $T_\rho$ &  \\
    $h_q$ &$ \mathbb{R}^{128} \rightarrow \mathbb{R}^M $& Quantization matrix encoder & Used in $\mathbf{A} = h_q(\mathbf{Q})\cdot h_a(\mathbf{z}')$  \\
    $\mathcal{S}$ & $    (\mathbf{z},\delta) \rightarrow \{ [s_i\cdot\mathbf{z}(\delta_i),\Delta\delta_i]^{j}_{i=1},\mathbf{c}\}$ & Sampling operator &  \\
    $U(\cdot)$ & $\mathbb{R}^{r_1H\times r_2W \times C}\rightarrow \mathbb{R}^{H\times W \times C}$ & Upsample operator &  \\
        $( \cdot )'$ &$\text{\textendash}$&  Variant of $(\cdot)$ that is structurally equivalent &  \\
        $\tilde{( \cdot )}$ & $\text{\textendash}$ &  $(\cdot)$ with distortions &  \\
        $\hat{( \cdot )}$ &$\text{\textendash}$&  Prediction to $(\cdot)$ with a neural network &  \\
    $HPF,\;LPF$ & $\text{\textendash}$ & High/Low-pass filters &  \\
    $\mathcal{L}_d,\mathcal{L}_r$ & $\text{\textendash}$ & Distortion / bitrate loss & For $\mathbf{Q}_\psi$  \\
    $\odot,\;\otimes$ &$\text{\textendash}$& Hadamard / Kronecker product & Element-/tensor-wise \\
    \bottomrule
    \end{tabular}
    \caption{Notation table for the main paper (Elements and Functions (calculations), respectively.}
    \label{tab:notation}
\end{table*}

\normalsize
\section{Introduction}
This supplementary material is intended to support the main paper.
We provide a notation table in \cref{tab:notation} to clarify the methodology and problem formulation. Additional background of JPEG compression codec is provided to define the problem and illustrate how the decoder works with a schematic overflow in \cref{fig:dec_process}.
Further experimental results are included.

\section{JPEG Preliminary}
In this section, we build upon \cref{sec:problemformulation} of the main paper and elaborate on the fundamentals of JPEG compression. We then characterize the compression artifacts of JPEG and provide further details of our proposed method.

\noindent{\bf Standard JPEG compression} \cref{fig:jpeg_conventional} hows the process of encoding and decoding an RGB image through the standard JPEG pipeline. During encoding, the JPEG algorithm decomposes an RGB image into its luminance and chroma components. The chroma components undergo a subsampling process using the nearest neighbor method. The chroma subsampling procedure is optional, with three distinct methods: reducing both dimensions by half (4:2:0), halving only the width (4:2:2), or maintaining both dimensions unchanged (4:4:4). Subsequently, $8\times8$ 2D-DCT transformed spectra are quantized using a predefined quantization matrix. 
During decoding, the stored quantization map ($\mathbf{Q}$) is multiplied with the quantized coefficients ($\mathbf{Y}'$) to recover the spectrum  ($\mathbf{\tilde{Y}}$). A 2D-IDCT is then applied, followed by resizing the chroma channels to the luminance resolution. The result is transformed back to the RGB domain. Each symbol is encoded into a bitstream via Huffman coding.

A backward-compatible neural codec with JPEG, including our JPNeO, is subject to several constraints within this process.
Most notably, JENO is applied as a pre-processing step before quantization. This constraint ensures that images compressed by JENO remain decodable into valid outputs using the standard JPEG decoder.
As noted by \citet{jdec2024han}, JDNO operates directly on spectral representations and be capable of reconstructing high-frequency components. To support arbitrary resizing of chroma components, a coordinate-based representation—such as implicit neural representations or neural operators—is applied.

\noindent{\bf JPEG Artifact Removal} JPEG compression loss mainly occurs during the encoding process and exhibits complex characteristics. In particular, the independent processing of blocks introduces noticeable visual discontinuities at block boundaries, commonly known as blocking artifacts. High-frequency components undergo more heavier quantization than low-frequencies, resulting in greater loss in high-frequency regions.
In addition, the chroma channels are resized during compression, color distortions are more severe than those in the luminance channel.

We hypothesize that JPEG distortions can be modeled as a differential equation, and our objective is to formulate and solve this equation accordingly. To realize this mechanism, we model the decoder as a neural operator.
In \cref{fig:dec_process}, a JPEG image \(f=\mathbf{L}_{a}u\) is stored as a {DCT spectrum}. We propose a cosine operator that predicts a continuous spectral representation and maps it to image features—serving two key purposes in bridging frequency and spatial domains.
The first is to infer high-frequency details, thereby improving the reconstruction of missing information.
The second is to bridge the spectral and image domains by translating frequency-domain information into spatial representations.
Then, a neural operator then approximates the inverse transform \(\mathbf{L}_{a}^{-1}\).

\section{Additional Experiments}

\begin{figure}[t]
\footnotesize
\centering
\includegraphics[trim= 0 0 0 0,clip,width = 3.2in]{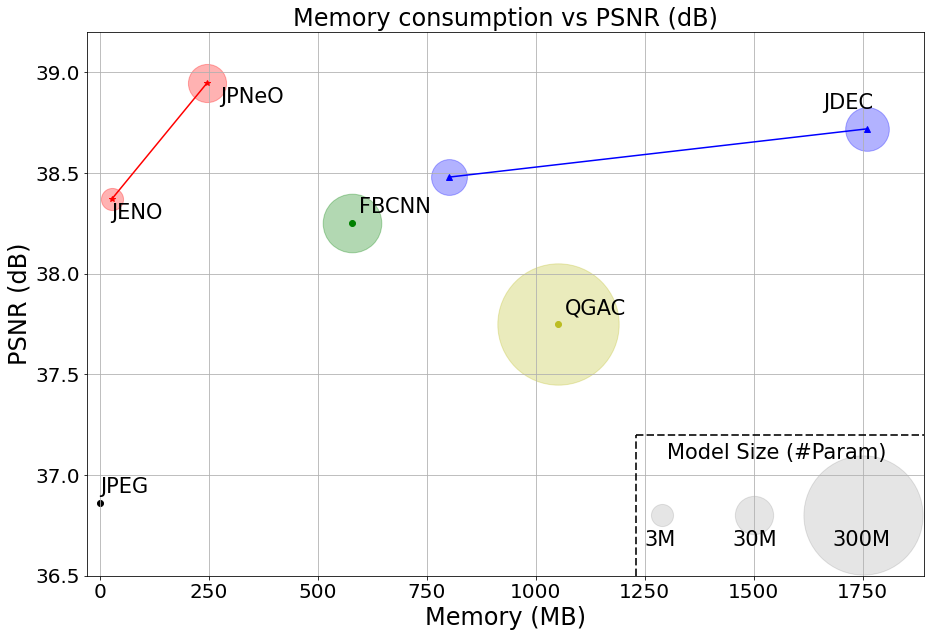}
\caption{\textbf{PSNR and memory consumption comparison} with other methods in LIVE-1 \cite{Live1}($q=90$, 4:2:0 subsampling).}
\label{fig:mem_psnr_comp}
\end{figure}

\noindent{\bf Memory Consumption} 
 Following the \cref{sec:discussuin} of the main paper, we report memory consumption and the number of parameters. We compare our JPNeO with QGAC \cite{qgac}, FBCNN \cite{fbcnn}, and JDEC \cite{jdec2024han}. The size of the input for the comparison is $560\times 560$. \cref{fig:mem_psnr_comp} shows the result of the comparison. We also report the memory usage of JENO. Notably, JENO surpasses the performance of FBCNN and QGAC while requiring only a minimal amount of memory.

\begin{figure*}[!t]
\footnotesize
\centering
\vspace{-15pt}
\hspace{-25pt}
\stackunder[2pt]{\tiny{}}{\includegraphics[trim=0 0 0 0,clip,width = 0.24\linewidth]{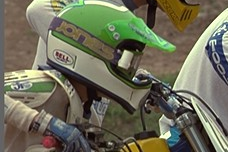}}
\stackunder[2pt]{\tiny{}}{\includegraphics[trim=0 0 0 0,clip,width = 0.24\linewidth]{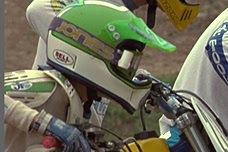}}
\stackunder[2pt]{\tiny{}}{\includegraphics[trim=0 0 0 0,clip,width = 0.24\linewidth]{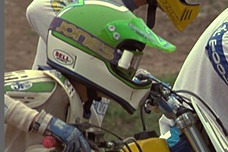}}
\stackunder[2pt]{\tiny{}}{\includegraphics[trim=0 0 0 0,clip,width = 0.24\linewidth]{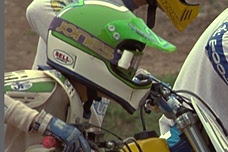}}

\hspace{-25pt}
\stackunder[2pt]{\includegraphics[trim=0 0 0 0,clip,width =  0.24\linewidth]{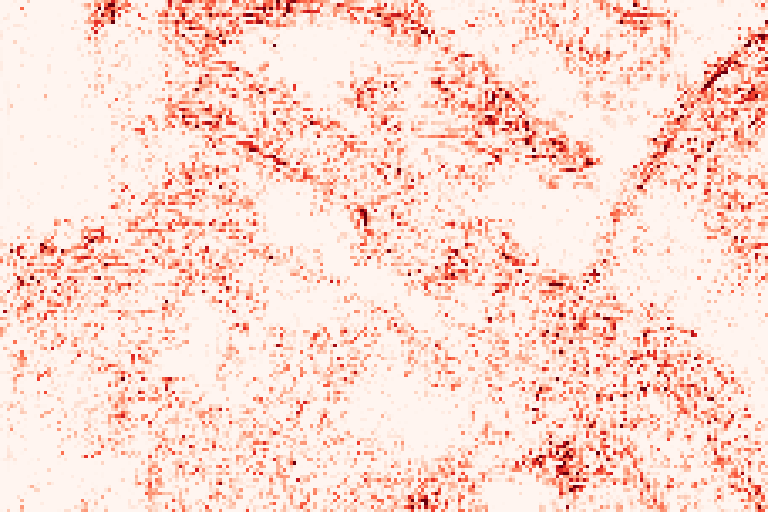}}{{JPEG+$\mathbf{Q}$}}
\stackunder[2pt]{\includegraphics[trim=0 0 0 0,clip,width = 0.24\linewidth]{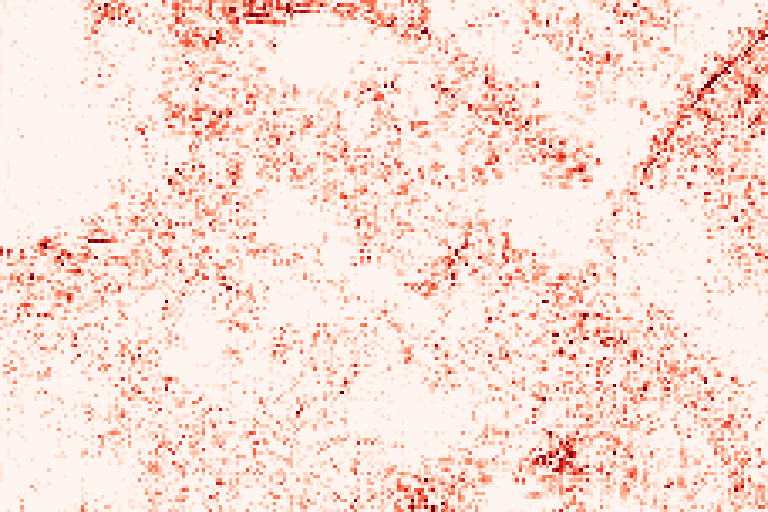}}{{JENO+$\mathbf{Q}$}}
\stackunder[2pt]{\includegraphics[trim=0 0 0 0,clip,width =  0.24\linewidth]{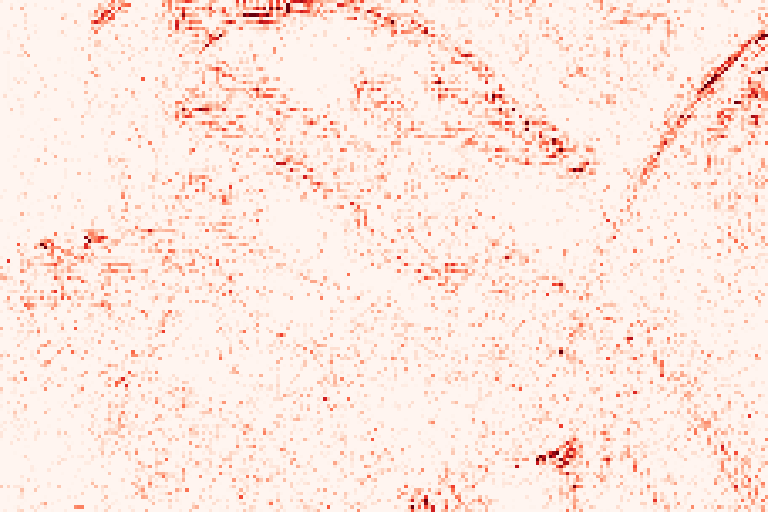}}{{JPEG+$\mathbf{Q}_\psi$}}
\stackunder[2pt]{\includegraphics[trim=0 0 0 0,clip,width = 0.24\linewidth]{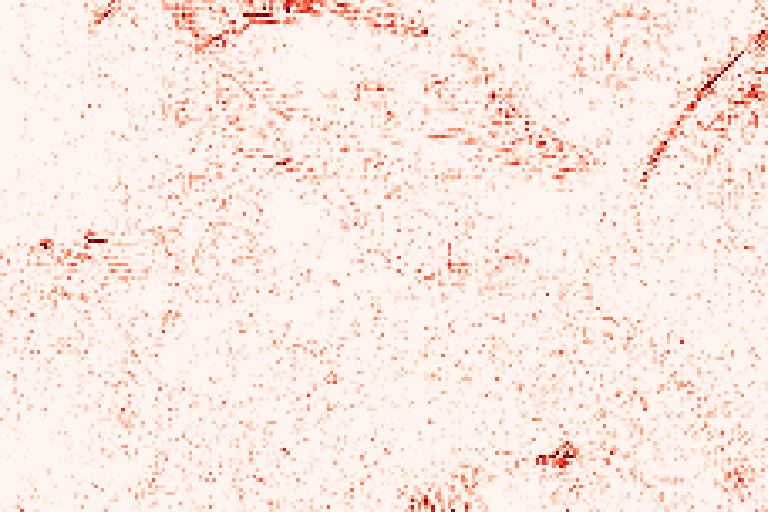}}{{JENO+$\mathbf{Q}_\psi$}}
\caption{Qualitative comparison images (top) and corresponding error maps (bottom) compressed by varying an encoder ($E_\text{JPEG}/E_\varphi$) and a quantization matrix ($\mathbf{Q}/\mathbf{Q}_\psi$).}
\label{fig:add_qual}
\vspace{-15pt}
\end{figure*}

\begin{wrapfigure}{r}{1.51in}
    \footnotesize
    \centering
    \includegraphics[trim={0 0 0 0},clip,width=1.5in]{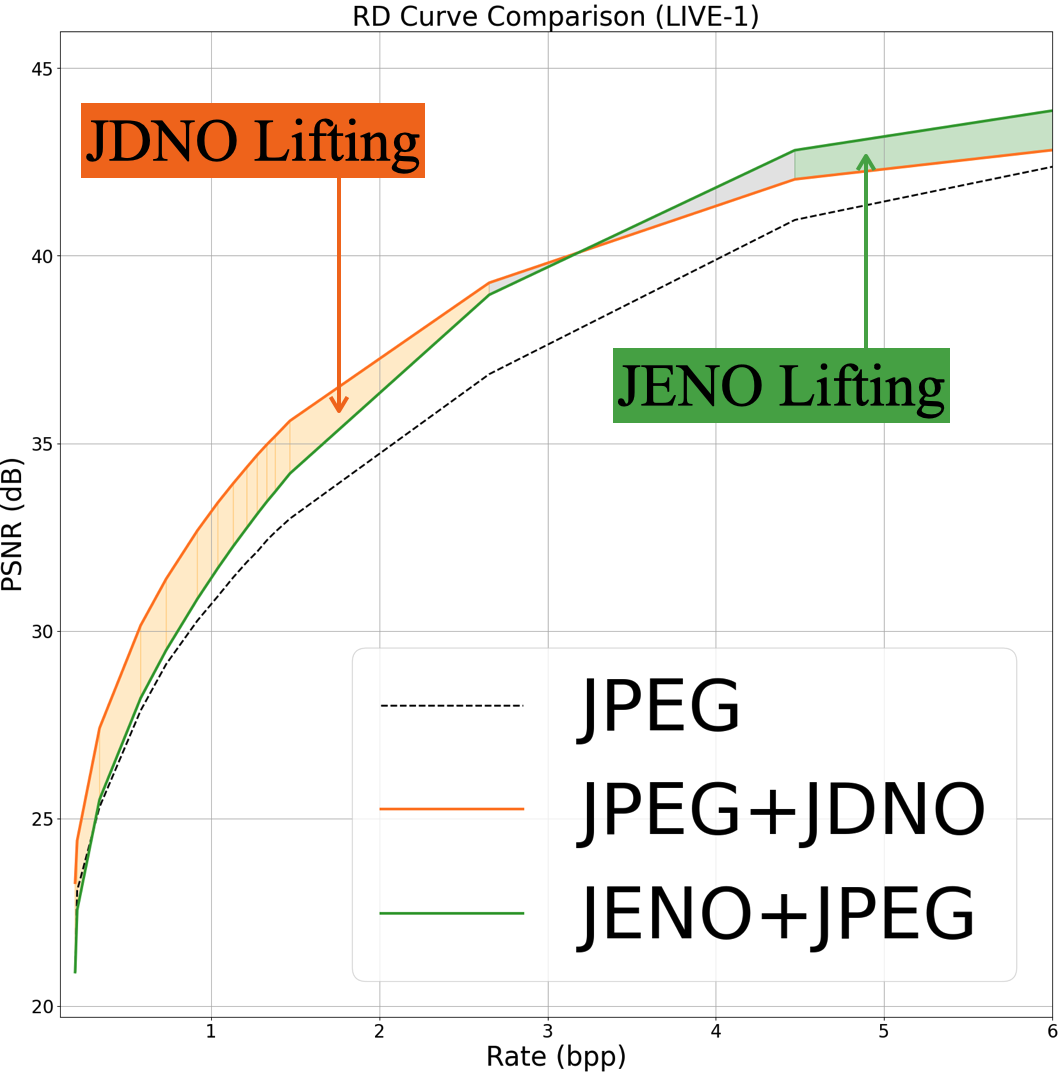}
    \vspace{-15pt}
    \caption{Improvements obtained by replacing $E_\text{JPEG}/D_\text{JPEG}$ to $E_{\varphi}/D_{\theta}$ compared to JPEG. The decoder contributes more significantly at low bpp levels, while the encoder becomes more influential as bpp increases.}
    \vspace{-20pt}
    \label{fig:jpneo_lifting}
\end{wrapfigure}

\noindent{\bf Roles of JDNO and JENO}
\cref{fig:jpneo_lifting} shows the results when each encoder and decoder is replaced with the JPNeO component, compared to the original JPEG. 
At low rates the dominant artifacts are from quantization and blocking; the decoder-side JDNO removes these artifacts over plain JPEG—see the shaded “JDNO lifting” band.
At higher rates the quantization artifact is minor, so quality is bounded by the encoder’s performance (“JENO lifting” band).
As a result, JDNO is crucial in the low-bit-rate region, while JENO dominates at high rates. 
\noindent
\cref{fig:jpneo_lifting} relates to \cref{sec:discussuin} and \cref{fig:mutual_info} of the main paper, where the increase in mutual information corresponds to improvements in PSNR.
Notably, JENO is trained in a distortion-blind manner; while its impact is limited in highly degraded regions, it enhances color fidelity in high-quality areas as noted in \cref{fig:qual_chroma_abl_self} of the main paper.
\cref{fig:add_qual} supports the observation by supplementing \cref{fig:qual_chroma_abl_self} in the main paper by highlighting the advantage of jointly using JENO and the learned quantization map.

\begin{table}[!t]
\centering
\setlength{\tabcolsep}{1.2pt}
\scriptsize
\begin{tabular}{c|c|c|c|c|c|c}
 
 \multirow{2}{*}{\tiny{bpp$|$PSNR}}& \multicolumn{2}{c|}{JPEG + JPEG} & \multicolumn{2}{c|}{JENO + JPEG} & \multicolumn{2}{c}{JPNeO}\\
 \cline{2-7}
 & low &high & low &high & low &high\\
\hline
$Q$     &0.358$|$25.29          &1.488$|$33.07             &0.358$|$25.31          &1.499$|$33.19 &0.358$|$\textbf{\textcolor{black}{27.77}}          &1.499$|$35.95  \\
$Q_\psi$&\textbf{0.335$|$25.50} &\textbf{\textcolor{black}{1.464}$|$34.03}    &\textbf{0.335$|$25.51} & \textbf{1.486$|$34.19}  &\textbf{0.335$|$\textcolor{black}{27.77}} & \textbf{1.486$|$36.86} \\
\end{tabular}
\caption{Quantitative comparison by varying a quantization matrix ($\mathbf{Q}/\mathbf{Q}_\psi$)}
\label{tab:sup_q}
\end{table}
 Further, in \cref{tab:sup_q} we provide a comparison between $Q/Q_\psi$. Since it is difficult to find a $Q_\psi$ that matches the bpp of each $Q$, we show image quality at the most similar bpp values. 

\begin{figure}[t]
    \footnotesize
    \centering
    \includegraphics[trim={0 0 0 0},clip,width=3.1in]{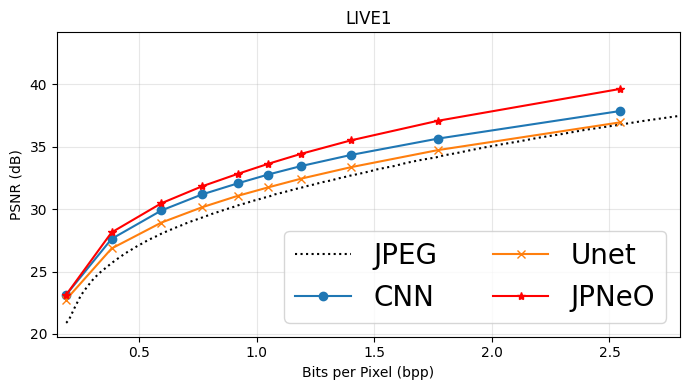}
    \caption{RD curve comparison with CNN \cite{Lim_2017_CVPR_Workshops} and U-Net\cite{fbcnn} architectures.}
    \label{fig:sup_ablation_rd}
\end{figure}

\noindent\textbf{Ablation Study} In \cref{tab:sup_ablation_quan} and \cref{fig:sup_ablation_rd}, we conducted ablation experiments using CNN and U-Net for the encoder and decoder. Specifically, we adopt the architecture introduced by \citet{Lim_2017_CVPR_Workshops} for CNN and \citet{fbcnn} for U-net. Our method consistently achieves better results than existing approaches. For a fair comparison, all networks were configured to have the same number of parameters. The results demonstrate that our JPNeO achieves greater efficiency under equal capacity.



\begin{table}[t]
\centering
\setlength{\tabcolsep}{1.2pt}
\scriptsize
\begin{tabular}{c|c|c|c|c||c|c|c|c|c}
\multirow{2}{*}{Dec.} &\multicolumn{2}{c|}{LIVE1 \cite{Live1}}&\multicolumn{2}{c||}{B500 \cite{B500dataset}}& \multirow{2}{*}{Enc.}&\multicolumn{2}{c|}{LIVE1 \cite{Live1}}&\multicolumn{2}{c}{B500 \cite{B500dataset}}\\
\cline{2-5}\cline{7-10}
&0& 40&0&40 & &90& 100&90&100 \\
\hline\hline
JDNO &\textcolor{red}{23.25} & \textcolor{red}{32.17}&\textcolor{red}{23.20}&\textcolor{red}{31.89} &JENO &\textcolor{red}{37.20}&\textcolor{red}{45.47}&\textcolor{red}{37.84}&\textcolor{red}{48.61}\\
\hline
\hline
CNN \cite{Lim_2017_CVPR_Workshops}& \textcolor{blue}{23.14}& \textcolor{blue}{32.06}&\textcolor{blue}{23.11}&\textcolor{blue}{31.83}&{CNN \cite{Lim_2017_CVPR_Workshops}}&\textcolor{blue}{37.17} & \textcolor{blue}{45.21}& \textcolor{blue}{37.71}&\textcolor{blue}{47.15}\\
{UNet \cite{fbcnn}}& 22.77& 31.05& 22.76&30.95&{UNet \cite{fbcnn}}& 37.13&44.87&37.67 &46.68 \\
\end{tabular}
\caption{Quantitative ablation study by replacing components of our JPNeO with CNN \cite{Lim_2017_CVPR_Workshops} and U-Net \cite{fbcnn} architectures.}
\label{tab:sup_ablation_quan}
\end{table}

\end{document}